\documentclass[preprint,amsmath,amssymb,superscriptaddress]{revtex4-1}
\usepackage[utf8]{inputenc}
\usepackage[T1]{fontenc}
\usepackage{hyperref}
\usepackage{graphicx}
\usepackage{bm}

\DeclareMathOperator{\tr}{tr}
\newcommand{\vphi}{\varphi}
\newcommand{\vphid}{\varphi^*}
\newcommand{\phid}{\phi^*}
\newcommand{\VPHI}{\bm{\varphi}}
\newcommand{\PHI}{\bm{\phi}}

\newcommand{\MR}{\mathbf{R}}
\newcommand{\MG}{\mathbf{G}}
\newcommand{\MGamma}{\bm{\Gamma}}
\newcommand{\Q}{\mathbf{q}}
\newcommand{\X}{\mathbf{x}}
\newcommand{\vP}{\mathbf{p}}

\newcommand{\dtau}{d\tau}
\newcommand{\ddX}{d^d\X}

\begin{document}

\title{Weakly-interacting Bose-Bose mixtures from the functional renormalisation group}

\author{Felipe Isaule}
\affiliation{School of Physics and Astronomy, University of Glasgow, Glasgow G12 8QQ, United Kingdom}
\affiliation{Departament de F\'isica Qu\`antica i Astrof\'isica, 
Facultat de F\'{\i}sica, and Institut de Ci\`encies del Cosmos (ICCUB), Universitat de Barcelona, 
Mart\'i i Franqu\`es 1, E–08028 Barcelona, Spain}
\author{Ivan Morera}
\affiliation{Departament de F\'isica Qu\`antica i Astrof\'isica, 
Facultat de F\'{\i}sica, and Institut de Ci\`encies del Cosmos (ICCUB), Universitat de Barcelona, 
Mart\'i i Franqu\`es 1, E–08028 Barcelona, Spain}

\date{\today}

\begin{abstract}

We provide a detailed presentation of the functional renormalisation group (FRG) approach for weakly-interacting Bose-Bose mixtures, including a complete discussion on the RG equations. To test this approach, we examine thermodynamic properties of balanced three-dimensional Bose-Bose gases at zero and finite temperatures and find a good agreement with related works. We also study ground-state energies of repulsive Bose polarons by examining mixtures in the limit of infinite population imbalance. Finally, we discuss future applications of the FRG to novel problems in Bose-Bose mixtures and related systems.
\end{abstract}


\maketitle

The experimental realisation of Bose-Einstein condensation (BEC) with cold alkali atoms~\cite{anderson_observation_1995,davis_bose-einstein_1995,bradley_evidence_1995} has greatly increased the interest in degenerate low-temperature quantum gases in the past two decades~\cite{pitaevskii_bose-einstein_2016}. Weakly-interacting one-component Bose gases~\cite{andersen_theory_2004} received significant theoretical attention after those first experiments. Even though one-component Bose gases have been theoretically studied for more than fifty years with perturbative approaches~\cite{bogoliubov_n_theory_1947,lee_eigenvalues_1957,lee_many-body_1957}, accurate descriptions were only achieved more recently with more robust approaches~\cite{prokofev_two-dimensional_2002,prokofev_weakly_2004,pilati_equation_2006,pilati_critical_2008,astrakharchik_equation_2009,astrakharchik_low-dimensional_2010,capogrosso-sansone_beliaev_2010}. Thanks to these developments, both experimental and theoretical efforts have shifted towards more sophisticated systems such as quantum mixtures~\cite{pitaevskii_bose-einstein_2016}.

In cold atom physics, quantum mixtures refer to gases with atoms in two or more internal states or with different atom species. The most prominent and maybe obvious example corresponds to two-component spin 1/2 Fermi gases such as BCS superfluids~\cite{strinati_bcsbec_2018}. In recent years two-component Bose gases, or Bose-Bose mixtures, have attracted significant attention. Because of the bosonic nature of both species, Bose-Bose mixtures can show a plethora of rich phenomena non-present in their fermionic counterparts. On top of having two separate superfluids at low temperatures, Bose-Bose mixtures can show phase separation~\cite{larsen_binary_1963,suthar_fluctuation-driven_2015}, droplet and liquid phases~\cite{petrov_quantum_2015}, spin drag~\cite{andreev_three-velocity_1975,fil_nondissipative_2005}, spin-orbit coupling~\cite{chen_spin-exchange-induced_2018}, amongst other phenomena. Experimentally, Bose-Bose mixtures were rapidly achieved after the first BEC experiments, both using atoms in two different spin states~\cite{myatt_production_1997,hall_dynamics_1998,hall_measurements_1998} and of two different atom species~\cite{modugno_two_2002}. However, only more recent experiments have been able to observe the novel droplet phases~\cite{cabrera_quantum_2018,semeghini_self-bound_2018,derrico_observation_2019}.

Perturbative calculations within Bogoliubov's theory can give a good description of homogeneous weakly-interacting Bose-Bose mixtures at low temperatures~\cite{larsen_binary_1963,petrov_quantum_2015,armaitis_hydrodynamic_2015,chiquillo_equation_2018,ota_beyond_2020,de_rosi_thermal_2021}. Improvements to perturbative calculations for finite-temperature mixtures have been provided by means of Popov's theory~\cite{konietin_2d_2018,ota_magnetic_2019,ota_thermodynamics_2020}, while strongly-interacting Bose-Bose liquids have been qualitatively explored with the introduction of pairing fields~\cite{hu_consistent_2020,hu_microscopic_2020-1}. Other employed approaches include Monte Carlo (MC) simulations ~\cite{petrov_ultradilute_2016,cikojevic_universality_2019}, Beliaev theory~\cite{utesov_effective_2018}, $N-$expansion~\cite{hryhorchak_large-n_2021}, and time-dependent Hartree-Fock theory~\cite{boudjemaa_quantum_2018,boudjemaa_many-body_2021}. Despite the recent efforts, Bose-Bose mixtures are by no means yet consistently described. There are still open questions, such as the phase diagram at finite temperatures and the onset of clustering in attractive mixtures.

Closely related is the problem of impurities immersed in dilute one-component Bose gases, which in the case of a single impurity is referred to as a Bose polaron. Experimentally, Bose polarons have been recently achieved with Bose-Bose~\cite{jorgensen_observation_2016} and Bose-Fermi~\cite{hu_bose_2016,yan_bose_2020} mixtures with huge population imbalances. Theoretically, Bose polarons have been intensively studied in the past few years with a variety of approaches~\cite{pena_ardila_impurity_2015,grusdt_renormalization_2015,levinsen_impurity_2015,christensen_quasiparticle_2015,camacho-guardian_landau_2018,yoshida_universality_2018,pastukhov_polaron_2018,pena_ardila_analyzing_2019,ichmoukhamedov_feynman_2019,hryhorchak_mean-field_2020}. In particular, many recent studies have focused on the role of multi-body correlations and the appearance of Efimov states~\cite{zinner_efimov_2013,levinsen_impurity_2015,sun_visualizing_2017}, as well as on their behaviour at finite temperatures~\cite{levinsen_finite-temperature_2017,guenther_bose_2018,pascual_quasiparticle_2021,field_fate_2020}. Aside from being interesting problems by themselves, the physics of Bose polarons can help elucidate analogous problems in Bose-Bose gases, such as the importance of multi-body correlations in attractive Bose-Bose mixtures and the formation of multi-body bound states. 

One promising approach to study Bose-Bose mixtures is the functional renormalisation group (FRG) based on the effective average action~\cite{wetterich_exact_1993,berges_non-perturbative_2002,dupuis_nonperturbative_2021}. The FRG is a field-theory technique where fluctuations, both quantum and thermal, are taken into account non-perturbatively by solving a RG equation, making it particularly suitable to study strongly-correlated systems and phase transitions. In the context of cold atom physics, the FRG has proved a powerful tool to study homogeneous cold quantum gases~\cite{boettcher_ultracold_2012}, including one-component Bose gases~\cite{floerchinger_functional_2008,floerchinger_superfluid_2009,floerchinger_nonperturbative_2009}, Fermi gases in the BCS-BEC crossover regime~\cite{diehl_functional_2010,scherer_functional_2011,boettcher_critical_2014}, Bose-Fermi mixtures~\cite{von_milczewski_functional_2021}, as well as related O(2)$-$models~\cite{grater_kosterlitz-thouless_1995,gersdorff_nonperturbative_2001,blaizot_non-perturbative_2005}. Similarly, strongly-correlated lattice systems have been studied with both fermionic~\cite{metzner_functional_2012} and bosonic~\cite{rancon_nonperturbative_2011,rancon_nonperturbative_2011-1} atoms. Recently, some of us have successfully applied the FRG to the study of both balanced repulsive Bose-Bose mixtures~\cite{isaule_functional_2021} and strongly-interacting Bose polarons~\cite{isaule_renormalization-group_2021}, both in homogeneous configurations at zero temperature. These results make the FRG a promising approach to give a consistent and unified description of Bose-Bose gases and Bose polarons.

In this work, we provide a detailed presentation of the FRG formalism for weakly-interacting Bose-Bose mixtures to motivate further FRG studies. To illustrate this approach, we examine thermodynamic properties of three-dimensional attractive and repulsive mixtures at both zero and finite temperatures, going beyond what was presented in our previous works~\cite{isaule_functional_2021,isaule_renormalization-group_2021}. Even though we do not explore new physics in this work, we narrow exciting new directions where the FRG could provide more robust descriptions than currently used approaches.

This article is organised as follow. In Sec.~\ref{sec:model} we present the microscopic model for Bose-Bose mixtures. In Sec.~\ref{sec:FRG} we provide a general presentation of the FRG approach, while in Sec.~\ref{sec:FRGBB} we detail the application of the FRG for Bose-Bose mixtures. We present results for balanced Bose-Bose gases in Sec.~\ref{sec:ResultsBoseBose} and for repulsive Bose polarons in Sec.~\ref{sec:ResultsBosePolaron}. Finally, in Sec.~\ref{sec:conclusions} we present conclusions and an outlook for future directions.
 
\section{Microscopic model for Bose-Bose mixtures}
\label{sec:model}

We consider a uniform gas with two species of bosons, $A$ and $B$, of masses $m_A$ and $m_B$ and chemical potentials $\mu_A$ and $\mu_B$, and that interact through weak short-range interactions. We consider that the interatomic interactions are dominated by $s$-wave scattering , and thus, we can model the interactions with contact two-body potentials of strength $g_{ab}$. The strengths $g_{ab}$ are connected to their respective $s-$wave scattering lengths $a_{ab}$ via the two-body $T-$matrices (see Sec.~\ref{sec:FRGBB;sub:IC}). Within the grand-canonical ensemble, such two-component gas is described by the Euclidean microscopic action~\cite{chiquillo_equation_2018}
\begin{equation}
\mathcal{S}[\VPHI]=\int_x
\Bigg[\sum_{a=A,B}\vphid_a(x)\left(\partial_\tau-\frac{\nabla^2}{2m_a}-\mu_a\right)\vphi_a(x)+\sum_{a,b=A,B}\frac{g_{ab}}{2}|\vphi_a(x)|^2|\vphi_b(x)|^2\Bigg]\,,
\label{sec:model;eq:S}
\end{equation}
where we use the notation $x=(\tau,\X)$, with $\tau=-it$ the imaginary time, and
\begin{equation}
    \int_x = \int_0^\beta \dtau \int \ddX\,,
\end{equation}
where $\beta=1/T$ is the inverse temperature. Note that even though in this work we focus on three-dimensional gases ($d=3$), the formalism is general. The microscopic action $\mathcal{S}$ is a functional of the fields $\VPHI=(\vphi_A,\vphid_A,\vphi_B,\vphid_B)$ which describe the two species of bosons\footnote{The complex fields $\varphi_a^*$ and $\varphi$ correspond to the field-theory formulation of the creation and annihilation operators, respectively. For a detailed formulation see Ref.~\cite{stoof_ultracold_2009}.}. The potential strengths $g_{AA}$ and $g_{BB}$ are associated with the intra-species interactions, whereas $g_{AB}=g_{BA}$ is associated with the inter-species interaction. Note that we use natural units $\hbar=k_B=1$.

The microscopic action defines the grand-canonical partition function as a \emph{path integral} over all configurations of the fields~\cite{stoof_ultracold_2009}
\begin{equation}
\mathcal{Z}[\VPHI]=\int \mathcal{D} \VPHI \, e^{-\mathcal{S}[\VPHI]}\,,
\label{sec:model;eq:Z}
\end{equation}
from which we can obtain the grand-canonical potential
\begin{equation}
 \Omega_G=-\beta^{-1}\ln \mathcal{Z}\,.
\end{equation}
Its differential is given by
\begin{equation}
 d\Omega_G=-Pd\mathcal{V}_d-SdT-\sum_{a=A,B}\langle N_a \rangle d\mu_a\,,
\label{sec:model;eq:dOmegaG}
\end{equation}
where $P$ is the pressure of the gas, $\mathcal{V}_d$ is the $d$-dimensional volume, $S$ is the entropy, and
$\langle N_a \rangle$ is the average number of particles of species $a$ in the thermodynamic limit. By differentiating $\Omega_G$ we can extract the thermodynamic properties of interest. In particular, the energy density is obtained from
\begin{equation}
    \epsilon = -P+sT+\sum_{a=A,B}n_a\mu_a\,,
    \label{sec:model;eq:epsilon}
\end{equation}
where $n_a=\langle N_a \rangle/\mathcal{V}_d$ is the atom density of species $a$ and $s=S/\mathcal{V}_d$ is the entropy density. The energy per particle is then obtained from $E/N=\epsilon/(n_A+n_B)$.

The microscopic action~(\ref{sec:model;eq:S}) shows a U$_A(1)\times$U$_B(1)$ symmetry associated with the conservation of particles of each species. The U$_a(1)$ symmetry is spontaneously broken at low temperatures in three dimensions and at zero temperature in two dimensions, signalling the condensation of species $a$. In this case, species $a$ develops a finite expectation value $\vphi_{a,0}=\langle\vphi_a\rangle$, which takes the role of the order parameter.

Throughout this article, we largely work in momentum space $q=(\omega_n,\Q)$, where $\omega_n=2\pi n T$ are the bosonic Matsubara frequencies. We employ the convention
\begin{equation}
    \VPHI(q)=\int_q e^{i(\omega_n\tau+\Q\cdot\X)}\VPHI(x)\,,
\end{equation}
where\footnote{Note that at zero temperature the Matsubara sum becomes the usual contour integral $T\sum_{n=-\infty}^\infty\to\int_{-\infty}^{\infty}\frac{d\omega}{2\pi}$.}
\begin{equation}
    \int_q =T\sum_{n=-\infty}^\infty\int \frac{d^d\Q}{(2\pi)^d}.
    \label{sec:model;eq:intq}
\end{equation}
Therefore, when species $a$ is condensed we have $\vphi_a(q)=\vphi_{a,0}(2\pi)^{d+1}\delta(q)$.

To compute the partition function~(\ref{sec:model;eq:Z}) we need to consider all the paths created by both quantum and thermal fluctuations (for detailed reviews on functional integration and field theory for quantum gases we refer to Refs.~\cite{popov_functional_1990,stoof_ultracold_2009,salasnich_zero-point_2016}). A commonly used approximation is to consider perturbative fluctuations around a mean-field (MF) solution~\cite{armaitis_hydrodynamic_2015,chiquillo_equation_2018}. A non-perturbative alternative is to employ the FRG. In the following, we present the FRG framework, which we then use to study Bose-Bose mixtures.

\section{The functional renormalisation group}
\label{sec:FRG}

Fluctuations are encoded in the paths that are followed by the microscopic action. As previously mentioned, these fluctuations can be taken into account with a perturbative expansion. An alternative is to work instead with an \emph{effective action} that already contains the effect of fluctuations. In the following, we present a short overview on how we can compute the effective action from the FRG. For detailed reviews on the FRG we refer to Refs.~\cite{berges_non-perturbative_2002,boettcher_ultracold_2012,dupuis_nonperturbative_2021}.

\subsection{The effective action}
\label{sec:FRG;sub:Gamma}

In order to define the effective action we first generalise the partition function (\ref{sec:model;eq:Z}) by introducing source fields $\bm{J}$,
\begin{equation}
\mathcal{Z}[\bm{J}]=e^{W[\bm{J}]}=\int \mathcal{D} \VPHI \, e^{-\mathcal{S}[\VPHI]+\int_x \bm{J}\cdot\VPHI}\,,
\end{equation}
which enables us to generate the $n$-point correlation functions from functional differentiation. Note that for a Bose-Bose mixture we have $\bm{J}=(J_A,J_A^*,J_B,J_B^*)$. However, the formalism is general.

We introduce \emph{classical fields} $\PHI$ from the 1-point function
\begin{equation}
    \PHI(x)=\langle\VPHI(x)\rangle=\frac{\delta W[\bm{J}
    ]}{\delta \bm{J}(x)}\,. 
\label{sec:FRG;sub:Gamma;eq:PHI}
\end{equation}
The effective action $\Gamma$ is then defined in terms of the classical fields by means of a Legendre transformation
\begin{equation}
    \Gamma[\PHI] = \left[-W[\bm{J}]+\int_x \bm{J}\cdot\PHI\right]_{\bm{J}=\bm{J}_0}\,, 
\label{sec:FRG;sub:Gamma;eq:Gamma}
\end{equation}
where $\bm{J}_0$ satisfies $\PHI=(\delta W[\bm{J}]/\delta\bm{J})_{\bm{J}_0}$. It is easy to check that $\Gamma$ is independent of $\bm{J}$. Here it is important to stress that because $\Gamma$ depends only on the classical fields $\PHI$, the effective action is a classical action that respects the stationary-action principle~\cite{berges_non-perturbative_2002}. 

The effective action is also referred to as the generator of the one-particle irreducible diagrams, and thus it encodes the solution of the theory. In particular, for vanishing source fields we have that the grand-canonical potential
\begin{equation}
    \Omega_G = T \,\Gamma[\PHI_0]\,,
    \label{sec:FRG;sub:Gamma;eq:OmegaG}
\end{equation}
where $(\delta\Gamma/\delta\PHI)_{\PHI_0}=0$. Therefore, we can easily extract the thermodynamic properties of a many-body system from the knowledge of $\Gamma$.

There are different ways to generate the effective action. For example, in a perturbative expansion we have
\begin{equation}
    \Gamma[\PHI] = \mathcal{S}[\PHI]+\frac{1}{2}\tr \ln \mathcal{S}^{(2)}[\PHI]+...\,.
\end{equation}
In the following, we show how to generate $\Gamma$ from the FRG.

\subsection{The FRG flow equation}
\label{sec:FRG;sub:FRG}

Within the FRG we introduce an infrared (IR) cutoff to the theory
\begin{equation}
    \mathcal{S}_k[\VPHI] = \mathcal{S}[\VPHI]+\Delta\mathcal{S}_k[\VPHI]\,,
    \label{sec:FRG;sub:FRG;eq:Sk}
\end{equation}
where $\Delta \mathcal{S}_k$ is defined as
\begin{equation}
    \Delta \mathcal{S}_k = \int_q \VPHI^\dag(q)\MR_k(q)\VPHI(q)\,.
    \label{sec:FRG;sub:FRG;eq:DSk}
\end{equation}
Here $k$ is a momentum scale and $\MR_k$ is referred to as the regulator function. The regulator can be chosen freely as long as its matrix elements $R_k$ satisfy
\begin{align}
R_k (q) & \xrightarrow{ k \to \infty }  \infty, \nonumber\\
R_k (q) & \xrightarrow{ k \to 0 \hspace{.2cm}}  0.
\end{align}
This ensures that for large scales $k$ the fluctuations in the path integral are suppressed, whereas for a vanishing cutoff at $k=0$ we have $\mathcal{S}_0=\mathcal{S}$. For convenience, we generally choose $R_k(q)\propto k^2$ for small $q$ so the scale $k$ acts as a mass term.

Analogously to the last section, we define the $k$-dependent partition function as
\begin{equation}
\mathcal{Z}_k[\bm{J}]=e^{W_k[\bm{J}]}=\int \mathcal{D} \VPHI \, e^{-\mathcal{S}_k[\VPHI]+\int_x \bm{J}\cdot\VPHI}\,,
\label{sec:FRG;sub:FRG;sec:Z}
\end{equation}
where $\mathcal{S}_k$ is defined in Eq.~(\ref{sec:FRG;sub:FRG;eq:Sk}). The $k$-dependent flowing effective action is then defined as
\begin{equation}
    \Gamma_k[\PHI] = \left[-W_k[\bm{J}]+\int_x \bm{J}\cdot\PHI\right]_{\bm{J}=\bm{J}_0}-\Delta\mathcal{S}_k[\PHI]\,.
\label{sec:FRG;sub:FRG;sec:Gammak}
\end{equation}
Note that $\Delta\mathcal{S}_k$ does not depend on $\bm{J}$. $\Gamma_k$ is the central object of the FRG. It connects the known microscopic physics in the ultraviolet (UV) for large $k$ with the macroscopic physics in the IR for small $k$. Indeed, for larges scales $k=\Lambda$, all fluctuations are suppressed by the cuttoff and so the effective action is simply the original microscopic action: $\Gamma_k=\mathcal{S}$. On the other hand, for $k\to 0$ all fluctuations are included, and thus $\Gamma_0$ corresponds to the physical effective action (\ref{sec:FRG;sub:Gamma;eq:Gamma}). We refer $k\to 0$ as the physical limit where we extract the solution of the model.

\begin{figure}[t]
\centering
\includegraphics[scale=0.6]{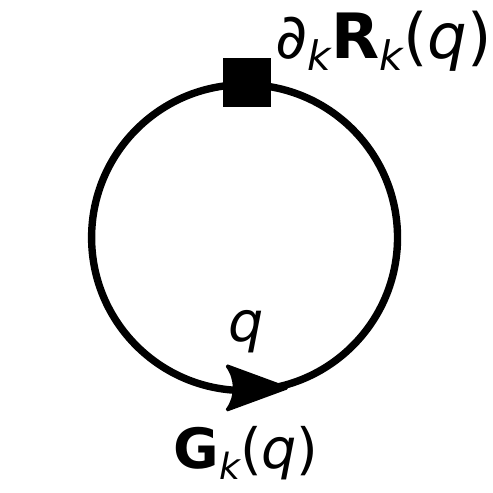}
\caption{Diagrammatic representation of the Wetterich equation~(\ref{sec:FRG;sub:FRG;eq:WettEq}). The solid line represents the propagator $\MG_k=(\MGamma^{(2)}_k+\MR_k)^{-1}$ and the square represents the insertion $\partial_k\MR_k$.}
    \label{sec:FRG;sub:FRG;fig:WetterichEq}
\end{figure}

To determine the \emph{flow} of $\Gamma_k$ as a function of $k$ we employ the Wetterich equation~\cite{wetterich_exact_1993}
\begin{equation}
    \partial_k \Gamma_k = \frac{1}{2}\int_q\tr[\partial_k\MR_k(\MGamma_k^{(2)}+\MR_k)^{-1}]\,,
    \label{sec:FRG;sub:FRG;eq:WettEq}
\end{equation}
where the matrix $\MGamma^{(2)}$ corresponds to the second functional derivative of $\Gamma_k$,
\begin{equation}
    \MGamma_k^{(2)}(q,-q)=\frac{\delta^2\Gamma_k[\PHI]}{\delta\PHI^\dag(-q)\PHI(q)}\,.
\end{equation}
The Wetterich equation dictates the flow of $\Gamma_k$ as a function of $k$. It has a one-loop structure where $\MG_k=(\MGamma^{(2)}_k+\MR_k)^{-1}$ is the propagator and $\partial_k \MR_k$ is an insertion (see Fig.~\ref{sec:FRG;sub:FRG;fig:WetterichEq}). By choosing a regulator $\MR_k$, and provided an initial condition $\Gamma_\Lambda=\mathcal{S}$, we can solve Eq.~(\ref{sec:FRG;sub:FRG;eq:WettEq}) from $k=\Lambda$ to  $k\to 0$. We provide a short derivation of the Wetterich equation in App.~\ref{app:WettEq}.

\begin{figure}[t]
\centering
\includegraphics[scale=0.75]{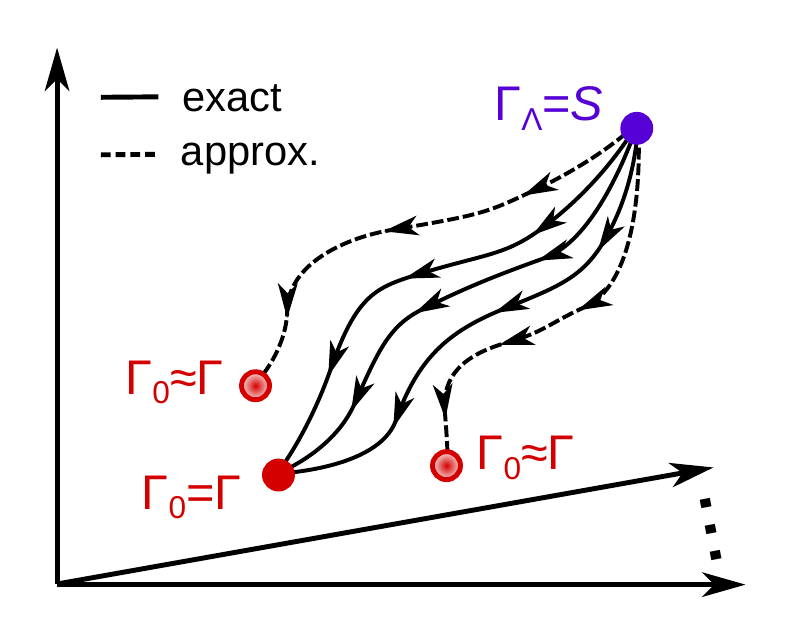}
\caption{Illustration of the RG flow of $\Gamma_k$ in parameter space~\cite{dupuis_nonperturbative_2021}. The solid curves represent exact flows with different regulator choices. The dashed curves represent flows with a truncated ansatz.}
    \label{sec:FRG;sub:FRG;fig:flow}
\end{figure}

We illustrate the FRG flow in Fig.~\ref{sec:FRG;sub:FRG;fig:flow}. Note that Eq.~(\ref{sec:FRG;sub:FRG;eq:WettEq}) is exact. Therefore, even though $\Gamma_k$ has distinct flows for different regulator choices, in principle we should obtain the same physical effective action for $k\to 0$ (solid lines). However, this picture is only correct if we solve the FRG flow exactly. In most applications, $\Gamma$ depends on an infinite number of couplings, and thus, we cannot solve the Wetterich equation exactly. In general, one proposes a truncated ansatz for $\Gamma_k$ which enables us to obtain an approximate solution for the effective action (dashed lines). Nevertheless, the Wetterich equation is non-perturbative and even simple approximations can provide accurate results for several problems. For more details see Refs.~\cite{litim_convergence_2002,litim_towards_2007}.

In the following, we propose an ansatz for the effective action of a Bose-Bose mixture based on a \emph{derivative expansion} (DE). Within a DE we expand $\Gamma_k$ in terms of the fields and their derivatives, always respecting the symmetries of the underlying theory. We truncate the expansion to a small number of $k$-dependent terms, and so the Wetterich equation becomes a set of coupled differential equations for the $k$-dependent couplings. These equations can be then solved numerically using standard methods. The DE has proved to be a robust approximation to study the thermodynamics of quantum gases and can even provide reasonable estimates for critical exponents~\cite{litim_derivative_2001}. However, in order to provide accurate descriptions of few-body physics and critical phenomena we need to employ a more sophisticated approximation, such as a vertex expansion. For more details see Ref.~\cite{berges_non-perturbative_2002,dupuis_nonperturbative_2021}.

\section{The FRG for Bose-Bose mixtures}
\label{sec:FRGBB}

To study a Bose-Bose mixture with the FRG we propose an ansatz based on a DE. In this work, we employ a generalisation of the ansatz used in Ref.~\cite{isaule_functional_2021} to imbalance  mixtures
\begin{equation}
\Gamma_k[\PHI]=\int_x
\Bigg[\sum_{a=A,B}\phid_a(x)\left(S_a\partial_\tau-\frac{Z_a}{2m_a}\nabla^2-V_a\partial^2_\tau\right)\phi_a(x)+U(\rho_A(x),\rho_B(x))\Bigg]\,,
\label{sec:FRGBB;eq:Gammak}
\end{equation}
where $\rho_a(x)=\phi^\dag_a(x)\phi_a(x)$, and $S_a$, $Z_a$ and $V_a$ are $k$-dependent renormalisation factors which we consider as field-independent.  Ansatz (\ref{sec:FRGBB;eq:Gammak}) of course builds upon the microscopic action (\ref{sec:model;eq:S}), containing the leading terms in momenta and fields. The terms and couplings not present in the original action~(\ref{sec:model;eq:S}) capture the effect of fluctuations during the RG flow. In particular, note that we add quadratic-frequency terms $V_a\partial^2_\tau$ which are not present in the microscopic theory. These are necessary to correctly describe the IR regime where the theory develops phonons with linear dispersion~\cite{floerchinger_functional_2008}. We truncate the $k$-dependent \emph{effective potential} $U$ up to fourth-order in the fields
\begin{equation}
    U(\rho_A(x),\rho_B(x))=-P+\sum_a u_a(\rho_a(x)-\rho_{a,0})+\sum_{a,b=A,B}\frac{\lambda_{ab}}{2}(\rho_a(x)-\rho_{a,0})(\rho_b(x)-\rho_{b,0})\,,
\label{sec:FRGBB;eq:U}
\end{equation}
where the minima of the potential $\rho_{a,0}=\langle\rho_a\rangle$ are the order parameters associated with each species. In addition to the renormalisation factors, the couplings $P$, $u_a$, and $\lambda_{ab}$, as well as $\rho_{a,0}$, all depend on $k$. Therefore, each $k$-dependent function has an associated flow equation. Note that $\lambda_{AB}=\lambda_{BA}$.

The effective potential evaluated at the minima satisfies
\begin{equation}
    \frac{\partial U}{\partial \rho_a}\Bigg|_{\rho_{0}}=0\,,
\end{equation}
for \emph{all} $k$ [see Eq.~(\ref{sec:FRG;sub:Gamma;eq:OmegaG})], where the subscript $\rho_0$ denotes that the expression is evaluated at $\rho_a=\rho_{a,0}$ ($a=A,B$). Therefore, when the $U_a(1)$ symmetry is broken we have that $\rho_{a,0}>0$ and $u_a=0$, whereas when the symmetry is unbroken we have that $\rho_{a,0}=0$ and $u_a>0$. In a Bose-Bose gas the symmetry is always broken in the UV (see Sec.~\ref{sec:FRGBB;sub:IC}) with a flowing $\rho_{a,0}>0$. However, at finite temperatures the flow can undergo a transition to a symmetric phase for $k<k^*$ where $u_a>0$ flows with $k$. The physical state of the system is dictated by the phase at $k=0$. Indeed, in a mixture gas we have that species $a$ is condensed if $\rho_{a,0}>0$ for $k\to 0$, whereas it is non-condensed otherwise\footnote{Note that in some problems the flows can undergo several transitions between symmetric and broken phases, such as in the pseudogap regime of Fermi gases~\cite{boettcher_critical_2014}. However, in a Bose gas if the flow is in the symmetric phase it remains there down to $k\to 0$.}. In particular, the phase transition occurs when $\rho_{a,0}$ flows to zero exactly at $k=0$. We examine some RG flows in Sec.~\ref{sec:FRGBB;sub:flows}.

The order parameters $\rho_{a,0}$ also correspond to the $k$-dependent condensate densities, giving their physical values for $k\to 0$. In addition, we can define the superfluid stiffness  $\rho_{a,s}=Z_a\rho_{a,0}$ from $\rho_{a,s}v_s^2/(2m_a)$~\cite{pitaevskii_bose-einstein_2016}, where $v_s=\nabla\theta_a$ is the superfluid velocity and $\theta_a$ is the phase of the condensate $\vphi=\rho_{a,0}^{1/2}e^{i\theta_a}$. In two and three dimensions $\rho_{a,s}$ corresponds to the superfluid density of species $a$~\cite{popov_functional_1990}, giving its physical value for $k\to 0$. In Bose gases we have that $\rho_{a,s}>\rho_{a,0}$. In particular, in quasicondensates gases we have that for $k\to 0$ the system is superfluid $\rho_{a,s}>0$, but the symmetry is unbroken $\rho_{a,0}=0$~\cite{dupuis_non-perturbative_2007}.

Another important ingredient in the model is the regulator choice. A commonly used choice for several problems is the optimised Litim regulator~\cite{litim_optimized_2001}. For Bose-Bose mixtures this regulator takes the form
\begin{equation}
    R^\text{opt}_{k,a}(\Q)=Z_a\frac{k^2-\Q^2}{2m_a}\Theta(k^2-\Q^2)\,,
    \label{sec:FRGBB;eq:Ropt}
\end{equation}
where $\Theta$ is the Heaviside step-function\footnote{Note that $R_{k,a}$ correspond to the diagonal elements in the regulator matrix $\MR_k$. The off-diagonal elements are zero.}. The optimised regulator enables us to perform the momentum integrals $\int_q$ analytically before solving the flow equations, greatly easing the numerical calculations. Even though this regulator is independent of $\omega_n$, and therefore it does not regulate the energies, it has proved successful in describing the thermodynamics of quantum gases~\cite{floerchinger_nonperturbative_2009}. For a complete review on optimised flows see Ref.~\cite{litim_critical_2002}. Another commonly used choice is an exponential regulator
\begin{equation}
    R^\text{exp}_{k,a}(\Q)=Z_a\frac{\Q^2/2m_a}{\exp(\Q^2/k^2)-1}\,,
    \label{sec:FRGBB;eq:Rexp}
\end{equation}
which has the benefit of having a smooth decay around $k$ instead of a sharp cutoff. In this work we employ only the optimised regulator~(\ref{sec:FRGBB;eq:Ropt}). For a detailed review on the dependence of the FRG flow on the regulator see Ref.~\cite{pawlowski_physics_2017}.

Finally, it is important to stress that, naturally, the accuracy of the results depends on the level of truncation of the ansatz. As we show later in this work, the current truncation gives accurate results at low temperatures. However, the accuracy decreases slightly as the temperature increases. Such accuracy can be improved by considering higher-order couplings in the fields and momenta. An analysis of the truncation level for one-component Bose gases can be found in Ref.~\cite{floerchinger_functional_2008}.

\subsection{Thermodynamics}
\label{sec:FRGBB;sub:thermo}

As previously discussed, the RG flow enables us to obtain the condensate and superfluid densities of the gas from $\rho_{a,0}$ and $Z_a\rho_{a,0}$.  To extract the thermodynamics we need to examine the grand-canonical potential $\Omega_G$.
From Eq.~(\ref{sec:FRG;sub:Gamma;eq:OmegaG}) we have that the physical ($k=0$) effective potential $U$ evaluated at $\rho_a=\rho_{a,0}$ gives the density of $\Omega_G$,
\begin{equation}
    U\Big|_{\rho_{a,0},k=0}=\Omega_G/\mathcal{V}_d\,,
\end{equation}
which enables us to extract thermodynamic quantities from differentiating the effective potential. From Eqs.~(\ref{sec:model;eq:dOmegaG}) and (\ref{sec:FRGBB;eq:U}) it is easy to see that the value of $P$ for $k\to 0$ simply corresponds to the physical pressure.

To obtain the atom densities we can differentiate $U$ with respect to $\mu_a$. 
Alternatively, an elegant and convenient way to extract the densities is to consider that the couplings depend on the chemical potentials (see Ref.~\cite{floerchinger_functional_2008} for details). If we consider that only the momentum-independent couplings depend on $\mu_a$, we can expand them as
\begin{align}
    -P\to& -P-\sum_{a=A,B}n_a (\tilde{\mu}_a-\mu_a)\,,\\
    u_a\to& u_a-\sum_{b=A,B}n^{(1)}_{a,b} (\tilde{\mu}_b-\mu_b)\,,\\
    \lambda_{ab}\to& \lambda_{ab}-\sum_{c=A,B}n^{(2)}_{ab,c} (\tilde{\mu}_c-\mu_c)\,,
\end{align}
where $\tilde{\mu}_a$ is a $k$-independent shift to the physical chemical potential $\mu_a$. Note that the superscripts (1) and (2) indicate that the couplings correspond to the first and second order terms in the expansion of $\rho_a$, respectively. We can identify $n_a$ as the $k$-dependent atom density of species $a$,
\begin{equation}
    n_a=\frac{\partial U}{\partial \tilde{\mu}_a}\Bigg|_{\rho_{0},\mu}\,,
\end{equation}
where we have evaluated at $\rho_a=\rho_{a,0}$ and $\tilde{\mu}_a=\mu_a$  ($a=A,B$). The physical values of the densities are obtained from the values of $n_a$ for $k\to 0$. Note that the densities flow with $k$, whereas the chemical potentials are fixed physical inputs. This is consistent with the fact that we work in the grand-canonical ensemble. Similarly, we can define the $k$-dependent entropy-density from\footnote{Note that because the complete finite-temperature behaviour of the flowing functions is encoded into the Matsubara sums, we do not need to consider an expansion around the temperature as with the chemical potential.}
\begin{equation}
    s=\frac{\partial U}{\partial T}\Bigg|_{\rho_0,\mu}\,,
\end{equation}
where the $T-$derivative is taken after performing the Matsubara sums.

As mentioned, the physical pressure is obtained from the value of $P$ for $k\to 0$. However, if we employ frequency-independent regulators the high energy modes of the zero-point function are not regulated, and so the Matsubara sums in the flow of the pressure do not converge properly~\cite{floerchinger_nonperturbative_2009}. To solve this issue we can set the renormalisation factors to their bare values in the UV~\cite{floerchinger_nonperturbative_2009}. Alternatively, we can calculate $P$ from the physical values of $n_a$ and $s$. From the Maxwell relations, the zero-temperature pressure can be obtained by integrating the zero-temperature densities
\begin{equation}
    P(\mu_a,T=0)=\sum_{a}\int_0^{\mu_a} d\mu'_a n_a(\mu'_a,T=0)\,,
\end{equation}
where we have used that the pressure in the vacuum is zero $P(\mu_a=0,T=0)=0$. The finite-temperature pressure is then obtained from integrating the entropy density
\begin{equation}
    P(\mu_a,T)=P(\mu_a,T=0)+\int_0^T dT' s(\mu_a,T')\,.
\end{equation}
By using the last two equations we can compute the pressure for any value of $\mu_a$ and $T$. 
The energy per particle can be then obtained from Eq.~(\ref{sec:model;eq:epsilon}).

\subsection{Propagator}
\label{sec:FRGBB;sub:G}

To solve the RG flow we need the $k$-dependent propagator $\MG_k=(\MGamma^{(2)}_k+\MR_k)^{-1}$. When we deal with a U(1) broken-symmetry it is convenient to introduce orthogonal real fields $\phi_{a,1}$ and $\phi_{a,2}$,
\begin{equation}
    \phi_a(x)=\frac{1}{\sqrt{2}}(\phi_{a,1}(x)+i\phi_{a,2}(x))\,.
\end{equation}
We fix the order parameters of both species at the same direction. Therefore, by imposing real background fields~\cite{floerchinger_functional_2008}, $\phi_{a,1}(x)=\sqrt{2\rho_a}$ and $\phi_{a,2}(x)=0$, the inverse propagator takes the form
\begin{equation}
    \MG^{-1}_k(q)=\begin{pmatrix}
    \MG^{-1}_{k,A}(q) & \bm{\Sigma}_{k,AB}\\
    \bm{\Sigma}_{k,BA} & \MG^{-1}_{k,B}(q)
    \end{pmatrix}\,,
    \label{sec:FRGBB;sub:G;eq:Ginv}
\end{equation}
where
\begin{gather}
    \MG^{-1}_{k,a}(q)=\begin{pmatrix}
    E_{a,1}(\Q;\rho_A,\rho_B)+V_B\omega_n^2 & S_a\omega_n\\
    -S_a\omega_n &  E_{a,2}(\Q;\rho_A,\rho_B)+V_a\omega_n^2 
    \end{pmatrix}\,,\\[0.5em]
    \bm{\Sigma}_{k,AB}=\bm{\Sigma}_{k,BA}=\begin{pmatrix}
    2\sqrt{\rho_{a}\rho_{b}}\lambda_{AB} & 0\\
    0 & 0
    \end{pmatrix}\,,
\end{gather}
and
\begin{align}
     E_{a,1}(\Q;\rho_A,\rho_B)=&E_{a,2}(\Q;\rho_A,\rho_B)+2\rho_a\partial^2_{\rho_a}U(\rho_A,\rho_B)\,,\label{sec:FRGBB;sub:G;eq:Ea1}\\
    E_{a,2}(\Q;\rho_A,\rho_B)=&Z_a\frac{\Q^2}{2m_a}+\partial_{\rho_a}U(\rho_A,\rho_B)+R_{k,a}(\Q)\,,\label{sec:FRGBB;sub:G;eq:Ea2}
\end{align}
where under our truncation $U$ is defined in Eq.~(\ref{sec:FRGBB;eq:U}). The propagator $\MG_k$ is obtained by inverting $\MG^{-1}$. This is then employed in the RG flow equations. 

Here we stress that because we fix both order parameters at the same direction, we do not take into the account the difference between the phases of the two condensates. Therefore, we do not describe the superfluid drag within the current approximation.

\subsection{Flow equations}
\label{sec:FRGBB;sub:floweqs}

The flow equations are obtained from projecting the flowing functions by functional differentiation. The flow of the effective potential $U$, including its couplings, is simply dictated by the flow of $\Gamma_k$: $\partial_k U=\partial_k \Gamma_k$ [see Eq.~(\ref{sec:FRGBB;eq:Gammak})]. Therefore, $\partial_k U$ is given by the one-loop Wetterich equation~(\ref{sec:FRG;sub:FRG;eq:WettEq}), where for our model $\MG_{k}$ is given by the inverse of Eq.~(\ref{sec:FRGBB;sub:G;eq:Ginv}) and $\partial_k \MR$ is obtained from differentiating our regulator choice~(\ref{sec:FRGBB;eq:Ropt}). The flow equations for the couplings in the expansion of $U$ are obtained by differentiating $\partial_k U$,
\begin{align}
    \partial_k u_a-\lambda_{aa}\partial_k\rho_{a,0}-\lambda_{AB}\partial_k\rho_{b,0}=&\frac{\partial}{\partial \rho_a}(\partial_k U)\Big|_{\rho_0,\mu}\,,& a\neq b  \label{sec:FRGBB;sub:floweqs;eq:vkua}\\
    \partial_k \lambda_{ab}=&\frac{\partial^2}{\partial \rho_a\partial\rho_b}(\partial_k U)\Big|_{\rho_0,\mu}\,,&\\
    \partial_k n_a-n^{(1)}_{a,a}\partial_k\rho_{a,0}-n^{(1)}_{b,a}\partial_k\rho_{b,0} =& -\frac{\partial}{\partial\tilde{\mu}_a}(\partial_k U)\Big|_{\rho_0,\mu}\,,& a\neq b\\
    \partial_k n^{(1)}_{a,b}-n^{(2)}_{aa,b}\partial_k\rho_{a,0}-n^{(2)}_{AB,b}\partial_k\rho_{b,0} =& -\frac{\partial^2}{\partial\rho_a\partial\tilde{\mu}_b}(\partial_k U)\Big|_{\rho_0,\mu}\,,& a\neq b\\
    \partial_k n^{(2)}_{ab,c} =& -\frac{\partial^3}{\partial\rho_a\partial\rho_b\partial\tilde{\mu}_c}(\partial_k U)\Big|_{\rho_0,\mu}\,,&\label{sec:FRGBB;sub:floweqs;eq:vknpp}
\end{align}
where all the equations are evaluated at $\rho_a=\rho_{a,0}$ and $\tilde{\mu}_a=\mu_a$ ($a=A,B$) after differentiating. We provide $\partial_k U$ in Eq.~(\ref{app:dkU;eq:dkU}). Note that Eq.~(\ref{sec:FRGBB;sub:floweqs;eq:vkua}) dictates the flow of both $u_a$ and $\rho_{a,0}$, which at a first sight it might suggest that we need an additional equation. However, in the broken phase we allow $\rho_{a,0}>0$ to flow while we simply set $\partial_k u_a=u_a=0$. On the other hand, in the symmetric phase we allow $u_a$ to flow while we set $\partial_k \rho_{a,0}=\rho_{a,0}=0$.

\begin{figure}[t]
\centering
\includegraphics[scale=0.4]{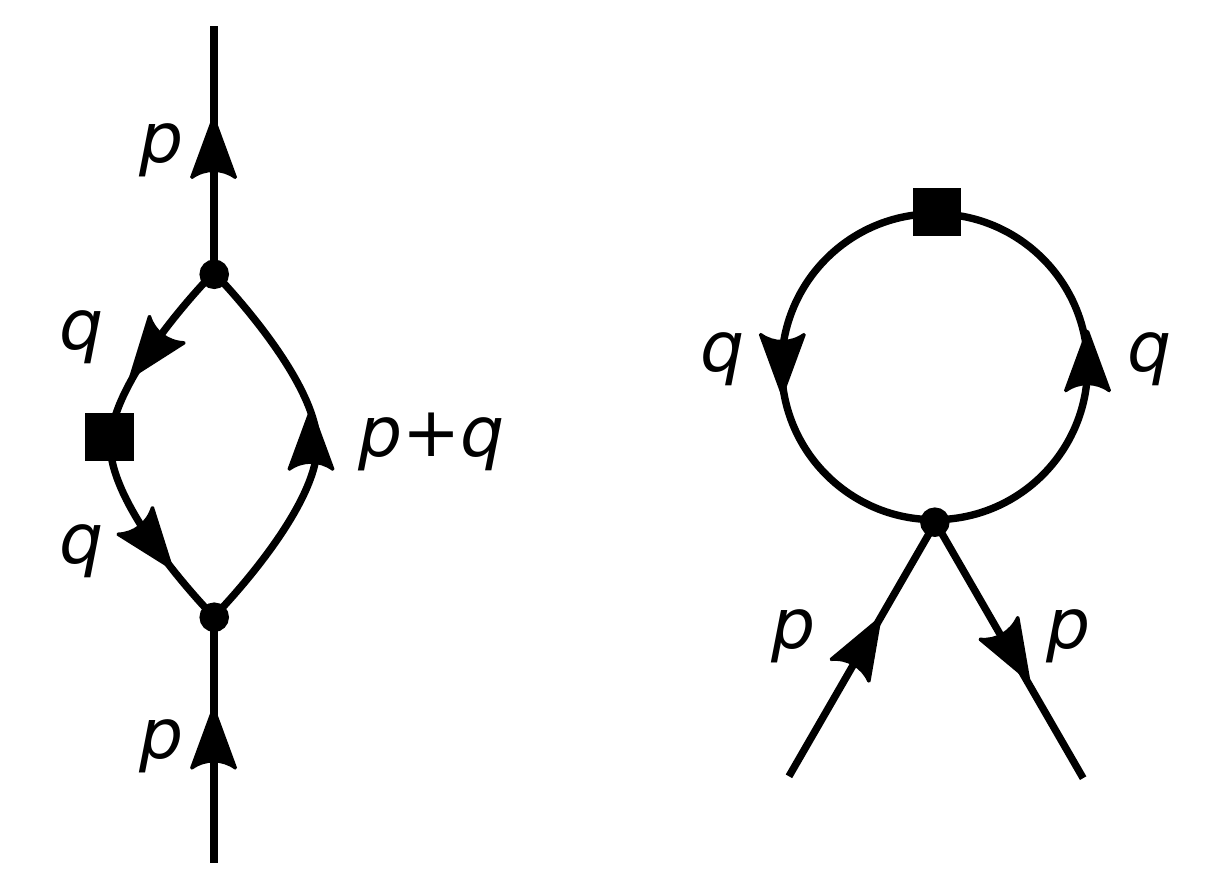}
\caption{Diagrams that contribute to the flow of the 2$-$point function $\MGamma_k^{(2)}$. The internal lines correspond to propagators $\MG_k$, while the black squares correspond to insertions of $\partial_k\MR_k$.}
\label{sec:FRGBB;sub:floweqs;fig:Gamma2}
\end{figure}

The flows of the renormalisation factors are dictated by the flow of the two-point function $\MGamma^{(2)}$. We have that [see Eq.~(\ref{sec:FRGBB;sub:G;eq:Ginv})]
\begin{align}
    \partial_k S_a = &\frac{\partial}{\partial \nu_n}(\partial_k \Gamma^{(2)}_{a_1,a_2})\Big|_{\rho_0,\mu,p=0}\,,\\
    \partial_k Z_a = &2m\frac{\partial}{\partial \vP^2}(\partial_k \Gamma^{(2)}_{a_2,a_2})\Big|_{\rho_0,\mu,p=0}\,,\\\partial_k V_a = &\frac{\partial}{\partial \nu^2_n}(\partial_k \Gamma^{(2)}_{a_2,a_2})\Big|_{\rho_0,\mu,p=0}\,,
\end{align}
where $p=(\nu_n,\vP)$ is an external momentum that is taken to zero after differentiating. The flow equation of $\MGamma^{(2)}$ is illustrated by the diagrams in Fig.~\ref{sec:FRGBB;sub:floweqs;fig:Gamma2}. We discuss the expressions for $\partial_k \MGamma^{(2)}$ in Eq.~(\ref{app:dkU;eq:dkGamma2}).

\subsection{Initial conditions and physical inputs}
\label{sec:FRGBB;sub:IC}

The RG flow is started in the UV at a high scale $k=\Lambda$. At this scale, we impose that $\Gamma_\Lambda=\mathcal{S}$ to obtain the MF-like initial conditions
\begin{gather}
    S_a(\Lambda)=Z_a(\Lambda)=1\,,\quad V_a(\Lambda)=0\,,\nonumber\\ n_{a}(\Lambda)=\rho_{a,0}(\Lambda)\,,\quad n^{(1)}_{a,b}(\Lambda)=\delta_{a,b}\,,\quad n^{(2)}_{ab,c}(\Lambda)=0\,.
    \label{sec:FRGBB;sub:IC;IC1}
\end{gather}
In the Bose-Bose gas phase we have
\begin{equation}
    u_a(\Lambda)=0\,,\quad\rho_{a,0}(\Lambda) = \dfrac{1}{1-\Delta(\Lambda)}\left[\dfrac{\mu_a}{\lambda_{aa}(\Lambda)}-\dfrac{\mu_b\lambda_{AB}(\Lambda)}{\lambda_{aa}(\Lambda)\lambda_{bb}(\Lambda)}\right]\,,\quad a\neq b
    \label{sec:FRGBB;sub:IC;ICBB}
\end{equation}
where $\frac{\mu_A}{\mu_B}>\frac{\lambda_{AB}(\Lambda)}{\lambda_{BB}(\Lambda)}$ and $\frac{\mu_B}{\mu_A}>\frac{\lambda_{AB}(\Lambda)}{\lambda_{AA}(\Lambda)}$, $\mu_A,\mu_B>0$, and
\begin{equation*}
    \Delta(\Lambda)=\frac{\lambda^2_{AB}(\Lambda)}{\lambda_{AA}(\Lambda)\lambda_{BB}(\Lambda)}\,.
\end{equation*}
Note that we choose $\Delta(\Lambda)<1$ to prevent phase-separation and collapse at the starting scale. In the Bose polaron phase~\cite{isaule_renormalization-group_2021}
\begin{gather}
    u_a(\Lambda)=0\,,\quad \rho_{a,0}(\Lambda)=\mu_a/\lambda_{aa}(\Lambda)\,,\nonumber\\
    u_b(\Lambda)=-\mu_b+\mu_a\frac{\lambda_{AB}(\Lambda)}{\lambda_{aa}(\Lambda)}\,,\quad \rho_{b,0}(\Lambda)=0\,,\qquad a\neq b
    \label{sec:FRGBB;sub:IC;ICBP}
\end{gather}
where $\mu_a>0$ and $\frac{\mu_b}{\mu_a}<\frac{\lambda_{AB}(\Lambda)}{\lambda_{aa}(\Lambda)}$. Here we can set $\lambda_{bb}=0$ to impose a single impurity\footnote{For the case of a single impurity the statistics is not important. Therefore, our approach is valid for both bosonic and fermionic impurities.}. Note that in this phase we can set $\lambda_{AB}>\lambda_{aa}$, in contrast to the Bose-Bose phase.
Finally, in the vacuum we have
\begin{equation}
    u_a(\Lambda)=-\mu_a\,,\quad \rho_{a,0}(\Lambda)=0\,,
\end{equation}
where $\mu_A,\mu_B<0$ are the single-atom energies. In the vacuum $n_a=0$ for all $k$, as expected.

We stress that to study the Bose-Bose liquid phase~\cite{petrov_quantum_2015,petrov_ultradilute_2016}, where $\lambda_{AB}<-(\lambda_{AA}\lambda_{BB})^{1/2}$ and $\mu_A,\mu_B<0$, we need a more sophisticated ansatz. Moreover, the MF-like initial conditions might not be appropriate to describe a liquid phase. Therefore, we leave the study of Bose-Bose liquids to future works.

The starting UV scale $\Lambda$ has to be chosen much larger than the physical many-body scales of the system where fluctuations do not play a role yet. Two of these scales correspond to the thermal scales $p_{a,T}=\sqrt{m_aT}$. The remaining two scales are set by the chemical potentials $\mu_A$ and $\mu_B$. In a balanced mixture ($m=m_A=m_B$, $\mu=\mu_A=\mu_B$, $g=g_{AA}=g_{BB}$), these are just the density and spin healing scales~\cite{chiquillo_equation_2018}
\begin{equation}
    p_{h,\pm} = \left(4m\mu\frac{g\pm g_{AB}}{g+g_{AB}}\right)^{1/2}\,,
\end{equation}
whereas for the Bose polaron phase these are given by $p_{h,a}=(4m_a\mu_ag_{aa})^{1/2}$ for $a=A,B$. For general expressions for imbalanced mixtures see Ref.~\cite{ota_thermodynamics_2020}. Note that in the previous expressions $g_{ab}$ just correspond to the two-body $T-$matrices, defined below.

\begin{figure}[t]
\centering
\includegraphics[scale=0.8]{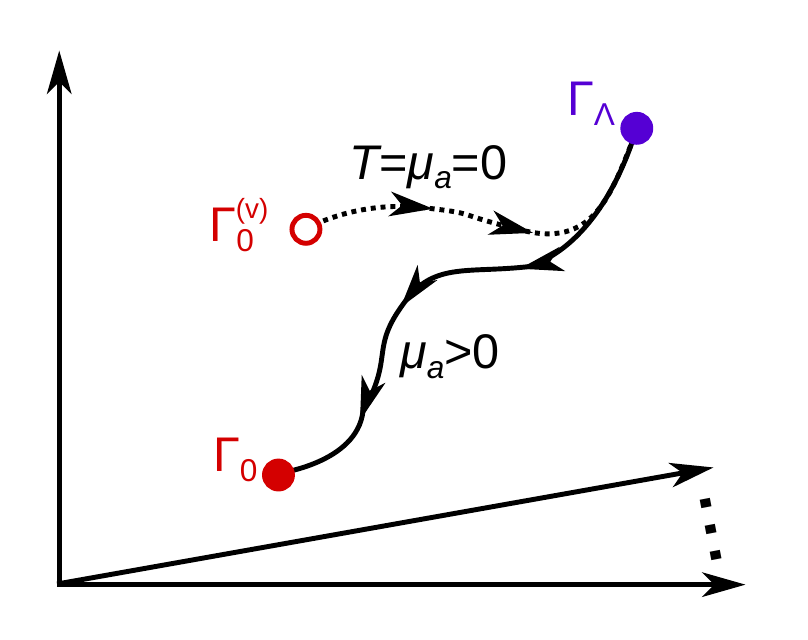}
\caption{Illustration of the RG flow of $\Gamma_k$ in theory space in vacuum $T=\mu_a=0$ (dashed line) and in-medium $\mu_a> 0$ (solid line). The flow is solved from the known vacuum solution $\Gamma_0^{(v)}$ at $k=0$ up to $k=\Lambda$. The latter is used as the initial condition in-medium.}
\label{sec:FRGBB;sub:IC;fig:flow}
\end{figure}

To connect the microscopic theory with physical scattering we need to renormalise the interaction couplings $\lambda_{ab}$~\cite{floerchinger_functional_2008}. Because at high scales many-body effects are not important, the RG flows in the UV in both the vacuum\footnote{In vacuum $\rho_{a,0}^{(v)}=n_{a,0}^{(v)}=0$ for all $k$ for both species.} and in-medium are indistinguishable. We use this property to generate initial conditions $\lambda_{ab}(\Lambda)$ from the known physics in vacuum. We illustrate this process in Fig.~\ref{sec:FRGBB;sub:IC;fig:flow}.

In vacuum ($T=\mu_a=0$) only the couplings $\lambda_{ab}$ flow with $k$, whereas all the other functions remain at their microscopic values~(\ref{sec:FRGBB;sub:IC;IC1}).
We impose that in vacuum  the physical couplings $\lambda_{ab}$ correspond to the known two-body $T-$matrices~\cite{floerchinger_functional_2008}
\begin{equation}
    \lambda^{(v)}_{ab}(k=0)=T_{ab}^{2B}\,.
    \label{sec:FRGBB;sub:IC;eq:lambdak0}
\end{equation}
In three dimensions these have the known form
\begin{equation}
     T_{ab}^{2B} = \frac{2\pi a_{ab}}{m_{ab}}\,,
    \label{sec:FRGBB;sub:IC;eq:T3D}
\end{equation}
where $m_{ab}=m_am_b/(m_a+m_b)$ is the reduced mass between particles $a$ and $b$, and $a_{ab}$ are the s$-$wave scattering lengths. Note that $m_{aa}=m_a/2$.

For frequency-independent regulators, we can perform the contour integrals analytically. We obtain that the flows of $\lambda_{ab}$ in vacuum are dictated by
\begin{equation}
     \partial_k \lambda^{(v)}_{ab} = \frac{\lambda^{(v)}_{ab}}{4}\int_\Q\left[\frac{\partial_k R_{k,a}(\Q)}{(\Q^2/2m_a+R_{k,a}(\Q))^2}+\frac{\partial_k R_{k,b}(\Q)}{(\Q^2/2m_b+R_{k,b}(\Q))^2}\right]\,,
\end{equation}
which can be solved in closed-form
\begin{align}
     \frac{1}{\lambda^{(v)}_{ab}}\Bigg|_{k=\Lambda}-\frac{1}{\lambda^{(v)}_{ab}}\Bigg|_{k=0} =\frac{1}{2}\int_\Q&\bigg[\frac{1}{\Q^2/2m_a+R_{\Lambda,a}(\Q)}-\frac{1}{\Q^2/2m_a+R_{0,a}(\Q)}\nonumber\\
     &+\frac{1}{\Q^2/2m_b+R_{\Lambda,b}(\Q)}-\frac{1}{\Q^2/2m_b+R_{0,b}(\Q)}\bigg]\,.
\end{align}
By imposing (\ref{sec:FRGBB;sub:IC;eq:lambdak0}) we can obtain $\lambda^{(v)}_{ab}(\Lambda)$, which we use as their initial conditions in-medium. For the optimised cutoff~(\ref{sec:FRGBB;eq:Ropt}) we obtain
\begin{equation}
    \lambda^\text{opt}_{ab}(\Lambda)=\left(\frac{m_{ab}}{2\pi a_{ab}}-\frac{2m_{ab}}{3\pi^2}\Lambda\right)^{-1}\,.
    \label{sec:FRGBB;sub:IC;lambdaIC3D}
\end{equation}
By using (\ref{sec:FRGBB;sub:IC;IC1}-\ref{sec:FRGBB;sub:IC;ICBP}) and (\ref{sec:FRGBB;sub:IC;lambdaIC3D}) the initial conditions of the RG flow are completely defined in terms of the physical inputs $m_a$, $\mu_a$, $a_{ab}$ and $T$. Initial conditions in lower dimensions can be obtained by using the corresponding $T-$matrices.

For a positive scattering length $a_{ab}>0$,  we see that Eq.~(\ref{sec:FRGBB;sub:IC;lambdaIC3D}) diverges at $\Lambda^*=3\pi/4a_{ab}\approx 2/a$, becoming negative for larger scales\footnote{The particular value of $\Lambda^*$ is regulator-dependent.}. In repulsive interactions where $\lambda_{ab}$ must be positive, the scattering length $a_{ab}>0$ sets a lower bound to the interaction range. Therefore, for the contact potential approximation to be valid, we must restrict the flow to distances larger than $\sim a_{ab}$, hence we must choose $\Lambda\lesssim a_{ab}^{-1}$, which also ensures weak interactions. Because the intra-species interactions are repulsive, these always set an upper limit to $\Lambda$. Therefore, we must choose a starting scale that satisfies $\Lambda\lesssim a_{ab}^{-1}$ (for $a_{ab}>0$), but that is also much larger than the physical scales of the system. Note that, in contrast, the scattering length of attractive interactions does not set a lower bound to the interaction range, and thus attractive inter-species interactions do not restrict $\Lambda$.

Finally, it is worth mentioning that an effective range $r_{a,eff}$, as defined from the two-body $T-$matrix, can be included within the FRG by considering an additional term in the ansatz with the form $Y_a\rho_a\nabla^2\rho_a$. The effective range would appear in the initial condition for $Y_a$~\cite{floerchinger_functional_2008,stoof_ultracold_2009}.

\subsection{RG flow examples}
\label{sec:FRGBB;sub:flows}

\begin{figure}[t]
\centering
\includegraphics[scale=1.3]{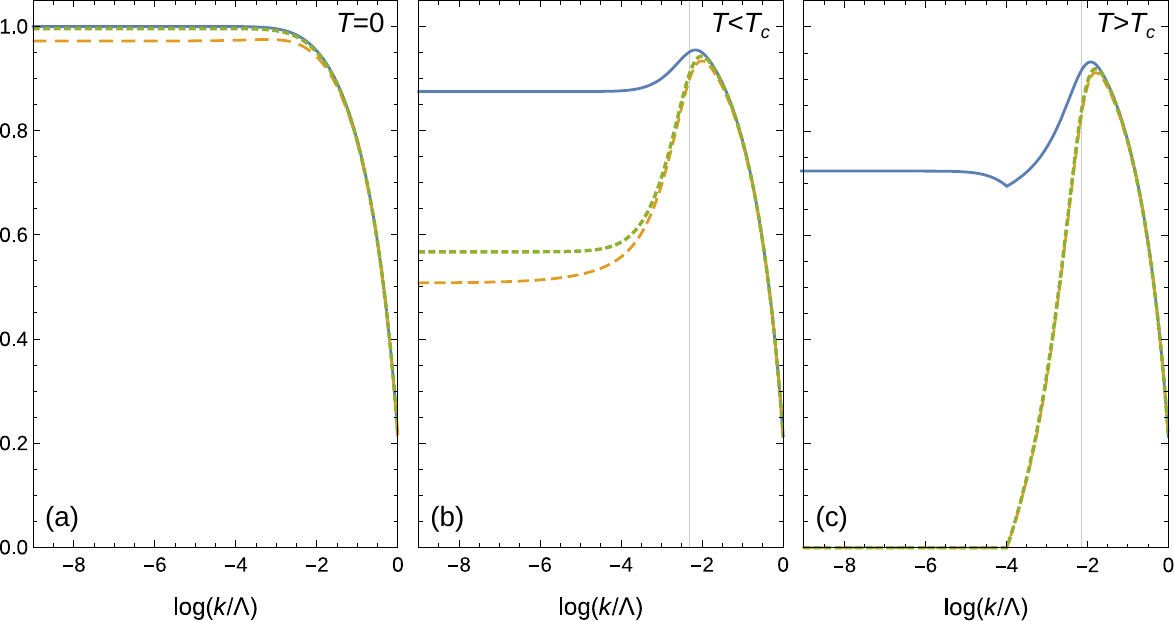}
\caption{Example flows of $n$ (blue, solid), $\rho_0$ (orange, dashed), and $\rho_s=Z \rho_0$ (green, dotted) as a function of $k$ for a balanced mixture  with a repulsive inter-species scattering length $a_{AB}=0.5a$ and the same chemical potential $\mu ma^2=0.1$. All the flows are normalised to the value of $n(k=0)$ at $T=0$. Panel (a) shows flows at zero temperature, panel (b) shows flows at a finite temperature $T<T_c$, and panel (c) shows flows at a finite temperature $T>T_c$, where $T_c$ is the critical temperature of the superfluid phase transition. The vertical lines correspond to the corresponding thermal scales $p_T=\sqrt{mT}$. }
\label{sec:FRGBB;sub:flows;fig:flows}
\end{figure}

To provide an example of how the couplings flow, in Fig.~\ref{sec:FRGBB;sub:flows;fig:flows} we show the flows of the condensate density $\rho_0$, the superfluid density $\rho_s=Z \rho_0$, and the atom density $n$ for a balanced ($m=m_A=m_B$, $\mu=\mu_A=\mu_B$, $a=a_{AA}=a_{BB}$) repulsive mixture ($a_{AB}/a>0$). Note that in a balanced mixture, we have $n=n_A=n_B$, $\rho_0=\rho_{A,0}=\rho_{B,0}$, and $Z=Z_A=Z_B$. We show flows at zero temperature ($T=0$), at a finite temperature below the superfluid critical temperature ($T<T_c$), and at a finite temperature in the normal phase ($T>T_c$). Solving one RG flow for a balanced mixture takes a few seconds for zero temperature in a modern laptop, whereas it takes at most a few minutes for finite temperatures.

At the starting scale $k=\Lambda$ (the right of each panel), all the densities are equal. This comes directly from the initial conditions (see the previous subsection). As $k$ is lowered, the RG flow incorporates fluctuations into the flowing functions, and thus, the densities flow to different values. We extract the physical values of the densities for the chosen input parameters ($a$, $a_{AB}$, $\mu$ and $T$) from the values of the densities for $k\to 0$ (the left of each panel). Note that we do not solve the flow down to $k=0$. Instead, one solves the flow only till the flows converge at a small scale $k$ $\emph{smaller}$ than the physical many-body scales. Also note that generally, one solves the flow in terms of the RG time $t=\ln(k/\Lambda)$ to better capture the coarse-graining at different scales.

At zero temperature both $\rho_s$ and $n$ have indistinguishable flows, and so $\rho_s=n$. This is consistent with the fact that in zero-temperature Bose gases all the bosons are superfluid. In contrast, the condensate density $\rho_0$ flows to a slightly smaller value, consistent with the known depletion of the condensate~\cite{lee_eigenvalues_1957,lee_many-body_1957}.

For $T<T_c$ we have that $n>\rho_s>\rho_c$ at $k\to 0$ due to the effect of thermal fluctuations. Indeed, around the thermal scale $p_T=\sqrt{mT}$ (vertical line) there is an important depletion of both the superfluid and the condensate. Nevertheless, both $\rho_s$ and $\rho_0$ flow to positive finite values for $k\to 0$, and so the gas is still condensed. In contrast, for $T>T_c$ both $\rho_s$ and $\rho_0$ flow to zero at a finite scale $k^*$ due to the stronger thermal fluctuations. As previously mentioned, from $k^*$ the coupling $u_a$ (not shown) starts to flow. Because $\rho_s=\rho_0=0$ for $k\to 0$, the gas is not condensed. On the other hand, the atom density $n_0$ still flows to a finite value, as expected.

Similar flows are obtained for other parameters. For detailed examples of the RG flows of the other couplings we refer to Refs.~\cite{floerchinger_functional_2008,isaule_functional_2021,isaule_renormalization-group_2021}.

\subsection{Limitations of the ansatz and outlook}

The level of truncated used here [Eqs.~(\ref{sec:FRGBB;eq:Gammak}) and (\ref{sec:FRGBB;eq:U})] is able to provide accurate results for Bose-Bose mixtures in two and three dimensions, particularly at low temperatures. However, it is important to briefly stress the limitations of the current truncation and how it can be improved in future works.

First, ansatz~(\ref{sec:FRGBB;eq:Gammak}) is not able to correctly describe quasicondensates where kinetic terms of all orders in the fields become relevant in the IR~\cite{dupuis_non-perturbative_2007}. This includes two-dimensional superfluids at finite temperatures and one-dimensional superfluids at zero temperature. This issue results in unstable flows where the superfluid density incorrectly decays for small $k$. Nevertheless, superfluid two-dimensional gases at finite temperatures can still be described because thermodynamic properties converge before these instabilities~\cite{floerchinger_superfluid_2009}. Moreover, the RG flow signals the BKT transition as quasi-fixed points~\cite{gersdorff_nonperturbative_2001}. On the other hand, in one dimension phase fluctuations have an even stronger effect, and thus, the RG flow becomes unstable before observables can be extracted~\cite{dupuis_non-perturbative_2007}. Therefore, the current ansatz is only applicable to two- and three-dimensional gases.

As previously mentioned, for Bose-Bose gases, another missing effect in the ansatz is that of the superfluid or spin drag~\cite{andreev_three-velocity_1975,fil_nondissipative_2005}. This corresponds to the interaction between the phases of the two condensates. The effect of the superfluid drag increases as the temperature increases, having an important impact on the phase transition in two dimension~\cite{karle_coupled_2019}. Therefore, the current ansatz has a limited application for finite-temperature two-dimensional Bose-Bose gases.

Finally, even though the current ansatz can describe attractive inter-species interactions for arbitrary scattering lengths, we can only expect it to give a good description for weak interactions. For stronger attractive interactions, the account of bound states becomes important. A commonly used approach to deal with strong attractive interactions is to introduce pairing fields via a Hubbard–Stratonovich transformation. Such approach is the standard way to describe the BCS-BEC crossover with the FRG~\cite{diehl_functional_2010,scherer_functional_2011}, and we recently used it to describe strongly-interacting attractive Bose polarons~\cite{isaule_renormalization-group_2021}. The description of Bose-Bose mixtures with strong attractive inter-species interactions, particularly the Bose-Bose liquid regime, is left to future work.

\section{Balanced Bose-Bose gases}
\label{sec:ResultsBoseBose}

We first examine the application of our formalism to balanced ($m=m_A=m_B$, $\mu=\mu_A=\mu_B$, $a=a_{AA}=a_{BB}$) three-dimensional Bose-Bose mixtures. In a balanced mixture the couplings associated with the different species are equal ($S_A=S_B$, $Z_A=Z_B$, etc), greatly reducing the number of flow equations. Moreover, the flow equations have a much simpler form. Therefore, the numerical calculations for balanced gases are much less demanding. 

\begin{figure}[t]
\centering
\includegraphics[scale=1.45]{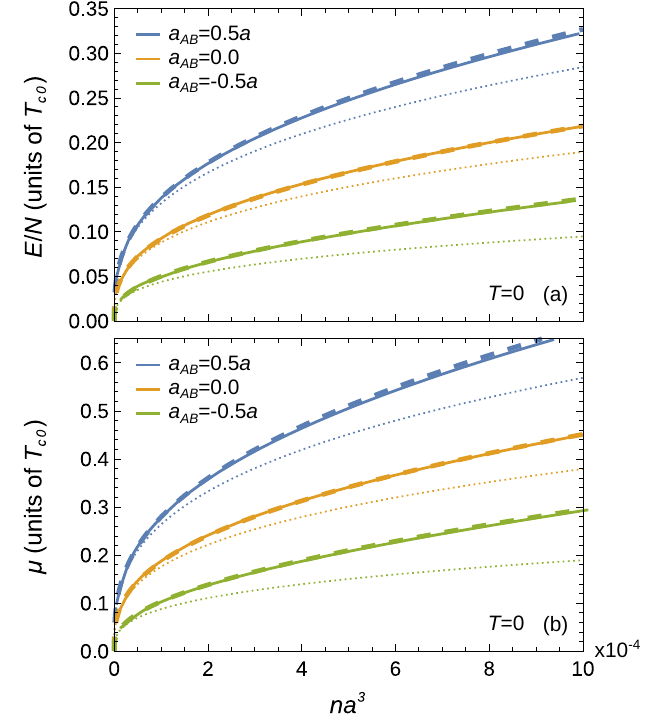}
\caption{Energy per particle $E/N$ (a) and chemical potential $\mu$  (b) at zero temperature for a balanced mixture as a function of the gas parameter $na^3$, where $n=n_A=n_B$ is the density of one species. The blue lines correspond to results for $a_{AB}=0.5a$, the orange lines to results for $a_{AB}=0$, and the green lines to results for $a_{AB}=-0.5a$. The thin dotted lines correspond to the MF solutions (\ref{sec:ResultsBoseBose;eq:EoN3D}-\ref{sec:ResultsBoseBose;eq:mu3D}), the thin dashed lines include the LHY corrections (\ref{sec:ResultsBoseBose;eq:EoN3D}-\ref{sec:ResultsBoseBose;eq:mu3D}), and the solid lines correspond to FRG calculations. }
\label{sec:ResultsBoseBose;fig:ThermoZT}
\end{figure}

In Fig.~\ref{sec:ResultsBoseBose;fig:ThermoZT} we show the energy per particle and chemical potential as a function of the gas parameter $na^{3}$, where $n$ is the atom density of one species, for a few choices of inter-species scattering lengths. We compare our results with the perturbative solutions for balanced mixtures~\cite{larsen_binary_1963}

\begin{align}
   \frac{E}{N}=&\frac{2\pi n}{m}\left(a+a_{AB}\right)+\frac{128\sqrt{\pi}}{15}\frac{n^{3/2}a^{5/2}}{m}f(a_{AB}/a)\,,\label{sec:ResultsBoseBose;eq:EoN3D}\\
   \mu=&\frac{4\pi n}{m}\left(a+a_{AB}\right)+\frac{64\sqrt{2\pi}}{3}\frac{n^{3/2}a^{5/2}}{m}f(a_{AB}/a)\,,\label{sec:ResultsBoseBose;eq:mu3D}  
\end{align}
where $f(x)=(1+x)^{5/2}+(1-x)^{5/2}$. The first terms on the right-hand-sides (RHS) correspond to the MF solutions, whereas the second terms to the LHY corrections. Note that we scale our results in terms of the critical temperature of an ideal Bose gas
\begin{equation}
    T_{c0}=\frac{2\pi}{m}(n/\zeta(3/2))^{2/3}\,.
\end{equation}

We obtain an excellent agreement with the LHY corrections, which are known to correctly describe three-dimensional mixtures at zero temperature for the employed gas parameters. We find that the FRG works well for both repulsive and attractive inter-species interactions. Indeed, we have found that we obtain similar agreements for other choices of $|a_{AB}/a|<1$. Note that the case $a_{AB}=0$ simply reduces the problem to two independent one-component Bose gases. For analogous comparisons for two-dimensional repulsive mixtures see Ref.~\cite{isaule_functional_2021}.

\begin{figure}[t!]
\centering
\includegraphics[scale=1.45]{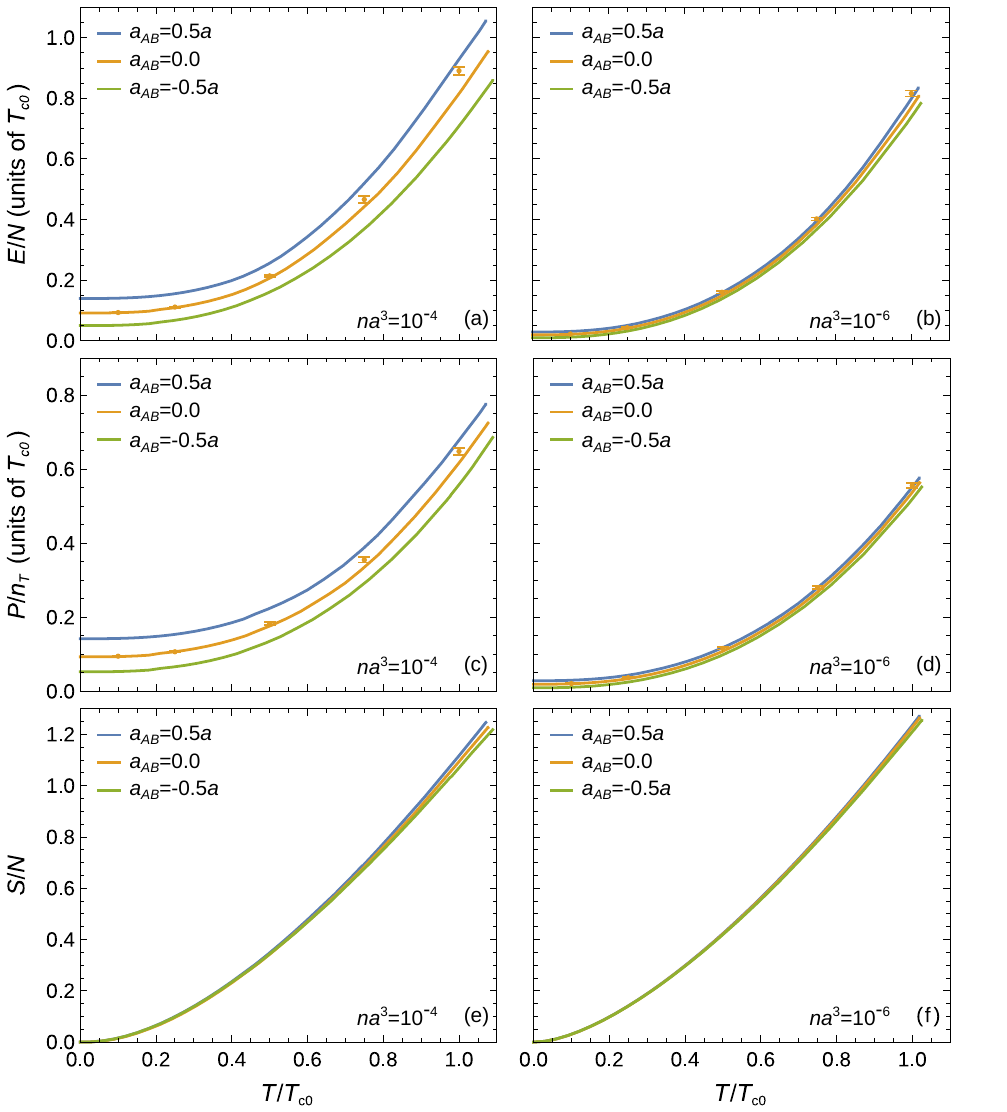}
\caption{Energy per particle $E/N$ (a,b), pressure over total density $P/n_T$ (c,d), and entropy per particle $S/N$ (e,f) for a balanced mixture as a function of the temperature for gas parameters $n a^3=10^{-4}$ (left panels) and $n a^3=10^{-6}$ (right panels), where $n=n_A=n_B$ is the density of one species and $n_T=n_A+n_B$. The blue lines correspond to results for $a_{AB}=0.5a$, the orange lines to results for $a_{AB}=0$, and the green lines to results for $a_{AB}=-0.5a$. The solid lines correspond to FRG calculations, while the squares are MC simulations for a one-component gas $a_{AB}=0$ from Ref.~\cite{pilati_equation_2006}.}
\label{sec:ResultsBoseBose;fig:ThermoFT}
\end{figure}

As explained previously, we can explore the finite-temperature regime by employing Matsubara sums. In Fig.~\ref{sec:ResultsBoseBose;fig:ThermoFT} we show the energy per particle, pressure and entropy at finite temperatures for two chosen gas parameters up to the superfluid phase transition. To benchmark our calculations, we compare with MC simulations for the one-component gas from Ref.~\cite{pilati_equation_2006}. Note that $S/N$ is dimensionless because we use $k_B=1$.

First, in the one-component gas limit ($a_{AB}=0$), at low temperatures ($T/T_{c0}\lesssim 0.75 $) there is an excellent agreement between the FRG calculations and MC simulations, whereas we obtain small deviations near the superfluid phase transition.
This is due to the simple ansatz used in this work. Our results can be improved by employing an ansatz that includes higher-order couplings and better regulator choices~\cite{blaizot_non-perturbative_2005,floerchinger_functional_2008,floerchinger_nonperturbative_2009}.

For finite choices of $a_{AB}$, we obtain that both the energy per particle and the pressure roughly maintain their effect from zero-temperature, with an almost constant displacement with respect to the case of $a_{AB}=0$ for different temperatures.  Nevertheless, these results are in reasonable agreement with what has been found with perturbative calculations~\cite{armaitis_hydrodynamic_2015}. On the other hand, the entropy is not much affected by the inter-species interaction, with only minor differences between the different curves. This behaviour agrees with what has been obtained by other works~\cite{armaitis_hydrodynamic_2015}.

Overall, the FRG can provide a good description of three-dimensional mixtures at finite temperatures, including the region around the phase transition. From the shown results, additional thermodynamic properties, such as compressibilities and sound velocities, can be easily extracted by using thermodynamic identities. Such complete examination of the thermodynamics of Bose-Bose mixtures is left for future work. Furthermore, even though here we have examined the thermodynamics for $T<T_c$, the normal phase is equally easy to access with the FRG. This will be examined in the future.

\section{Repulsive Bose polarons}
\label{sec:ResultsBosePolaron}

We now turn our attention to the limiting case of infinite population imbalance, where one species has a macroscopic finite density $n_B=n>0$, whereas the other species forms a gas of impurities with infinitesimal density $n_I=0$.  As explained in Sec.~\ref{sec:FRGBB;sub:IC}, this corresponds to the Bose polaron problem. In the following, we refer to the bosons in the Bose bath with the subscript $B$ and to the impurity with the subscript $I$. Moreover, here we focus only on the repulsive branch of the Bose polaron, where the boson-impurity interaction is repulsive and $a_{BI}>0$.

One of the most relevant properties of a polaron is its energy. This corresponds to the energy needed to add an impurity to the Bose bath. In our model, this energy corresponds to the impurity chemical potential $\mu_I$ [see Eq.~(\ref{sec:model;eq:S})]. However, $\mu_I$ is the physical ground-state energy only if it is the pole of the impurity propagator or, equivalently, the spectral function~\cite{rath_field-theoretical_2013}. In our formalism, the condition $\MG^{-1}(0)_{k=0,I}=0$ [Eq.~(\ref{sec:FRGBB;sub:G;eq:Ginv})] means that for $k\to 0$ the coupling $u_I$ must vanish\footnote{Note that for Bose polarons $\rho_{0,I}=0$, so $u_I$ flows instead.}. Therefore, to find the physical polaron energy $E_I=\mu_I$, we must find the choice of $\mu_I$ that gives a vanishing $u_I$ for a chosen set of parameters ($a_{BB}$, $a_{BI}$, $\mu_B$ and $T$), which can be done self-consistently. For a more detailed discussion we refer to Ref.~\cite{isaule_renormalization-group_2021}.

\begin{figure}[t]
\centering
\includegraphics[scale=1.3]{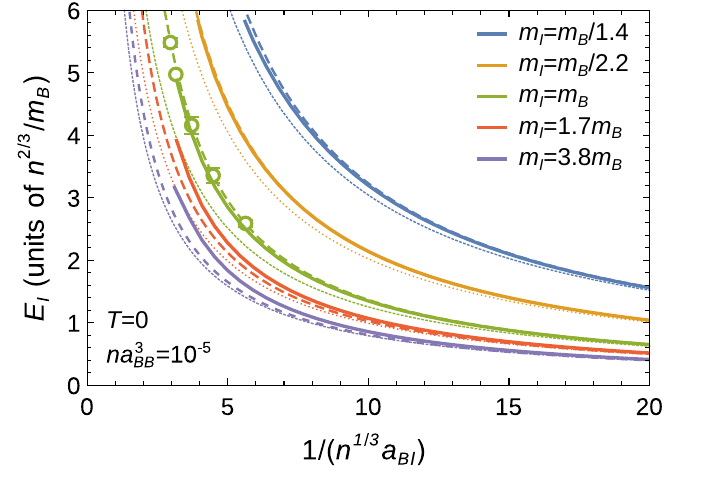}
\caption{Polaron energy $E_I$ at zero temperature as a function of $(n^{1/3}a_{BI})^{-1}$ for $n a_{BB}^3=10^{-5}$, where $n$ is the density of the bath. The different colours correspond to energies of impurities with different masses. The thin dotted lines correspond to the MF solution~(\ref{sec:ResultsBosePolaron;EI_3D}), the thin dashed lines include the first perturbative correction~(\ref{sec:ResultsBosePolaron;EI_3D}), the solid lines correspond to FRG calculations, and the circles correspond to MC simulations for $m_I=m_B$ from Ref.~\cite{pena_ardila_impurity_2015}.}
\label{sec:ResultsBosePolaron;fig:EIZT}
\end{figure}

We show polaron energies at zero temperature for different impurity masses\footnote{Bose polarons with equal masses $m_B=m_I$ have been achieved experimentally with  $^{39}K$ atoms in two different spin states~\cite{jorgensen_observation_2016}. The mass imbalance $m_I=m_B/2.2$ has been achieved with $^{40}$K impurities in a gas of $^{87}$Rb atoms~\cite{hu_bose_2016}, whereas $m_I=1.7m_B$ has been achieved with $^{40}K$ impurities in a gas of $^{23}$Na atoms~\cite{yan_bose_2020}. The imbalance $m_I=m_B/1.4$ and $m_I=3.8m_B$ could be achieved in the future with $^{23}$Na and $^{87}$Rb atoms.} and fixed gas parameter in Fig.~\ref{sec:ResultsBosePolaron;fig:EIZT}. We compare with MC simulations for $m_I=m_B$ from Ref.~\cite{pena_ardila_impurity_2015} and with the perturbative solution~\cite{christensen_quasiparticle_2015}

\begin{equation}
    E_I=\frac{2\pi a_{BI}n}{m_R}\left[1+\frac{24}{3\sqrt{\pi}}\frac{m_R}{m_I}\sqrt{n a_{BB}^3}\frac{a_{BI}}{a_{BB}}I(\gamma)\right]\,,
    \label{sec:ResultsBosePolaron;EI_3D}
\end{equation}
where $n=n_B$ is the density of the Bose bath, $\gamma=m_B/m_I$, and
\begin{equation*}
    I(\gamma)=\frac{1+\gamma}{\gamma}\int_0^\infty dk\left[1-\frac{(1+\gamma)k^2}{\sqrt{1+k^2}(\sqrt{1+k^2}+\gamma k)}\right]\,.
\end{equation*}
Note that for equal masses $m_B=m_I$, this takes the value $I(1)=8/3$. The first term on the RHS of Eq.~(\ref{sec:ResultsBosePolaron;EI_3D}) is the MF solution, whereas the second term is a LHY-type correction.

As reported previously in Ref.~\cite{isaule_renormalization-group_2021} for equal masses, the FRG calculations give a good agreement with the perturbative corrections and the MC simulations. In contrast, there are some deviations between the FRG and the perturbative results for heavier impurities $m_I>m_B$. Results for repulsive two-dimensional Bose polarons~\cite{isaule_renormalization-group_2021} suggest that the FRG performs better than perturbative solutions at the current truncation level. However, it is still difficult to give a conclusive conclusion without having simulations to compare with. Therefore, it is necessary to perform FRG calculations with additional higher-order couplings to check the convergence of our results. 

\begin{figure}[t]
\centering
\includegraphics[scale=1.5]{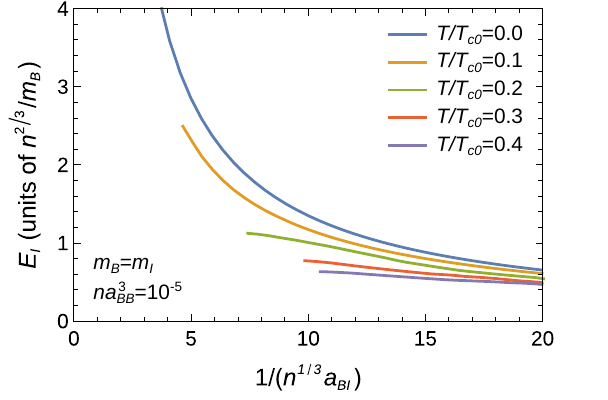}
\caption{Polaron energy $E_I$ obtained from the FRG as a function of $(n^{1/3}a_{BI})^{-1}$ for $m_B=m_I$ and gas parameter $n a_{BB}^3=10^{-5}$, where $n$ is the density of the bath. The different curves correspond to calculations at different temperatures $T$.}
\label{sec:ResultsBosePolaron;fig:EIFT}
\end{figure}

In Fig.~\ref{sec:ResultsBosePolaron;fig:EIFT} we now show polaron energies at finite temperatures for $m_B=m_I$. We observe that the energy decreases as the temperature increases, consistent with the trend predicted by recent MC simulations for similar gas parameters~\cite{pascual_quasiparticle_2021}. Therefore, the FRG seems to give, at least, a good qualitative description of finite-temperature polarons.

It is important to note that we present fewer results with increasing temperature. As discussed in Sec.~\ref{sec:FRGBB;sub:IC}, the starting scale $\Lambda$ must be chosen much larger than the many-body scales associated to $\mu_B$ and $T$, but also it must satisfy $\Lambda\lesssim \min(a^{-1}_{BB},a^{-1}_{BI})$. In the regime studied in Figs.~\ref{sec:ResultsBosePolaron;fig:EIZT} and ~\ref{sec:ResultsBosePolaron;fig:EIFT} we have that $a_{BI}\gg a_{BB}$. Therefore, $\Lambda$ is restricted in the UV by $a_{BI}$ only. At zero temperature, the many-body restriction set by $\mu_B$ means that we can only explore $(n^{1/3}a_{BI})^{-1}\gtrsim 5$. However, finite temperatures increasingly restrict $\Lambda$ to higher values, decreasing the region of $a_{BI}$ accessible. Therefore, with our current FRG formulation we can study relatively strong boson-impurity interactions at low temperatures only. 

Despite the mentioned shortcoming, the obtained results for finite temperatures makes the FRG a promising tool to study attractive Bose polarons at finite temperatures. As discussed in Sec.~\ref{sec:FRGBB;sub:IC}, in attractive interactions the scattering length does not set an upper bound to $\Lambda$. Therefore, by extending the pairing-fields approach for attractive Bose polarons employed in our previous work~\cite{isaule_renormalization-group_2021} to finite temperatures, we could study resonant Bose polarons for a wide range of temperatures.

\section{Conclusions} 
\label{sec:conclusions}

In this work, we give a detailed presentation of the FRG framework for weakly-interacting Bose-Bose mixtures by generalising previous works on balanced repulsive mixtures and Bose polarons. We apply this approach to study three-dimensional mixtures with both attractive and repulsive inter-species interactions, and also to study repulsive Bose polarons.

We find that the FRG gives a good description of three-dimensional mixtures at zero and finite temperatures in the cases studied. At finite temperatures, we expect deviations from exact calculations, which is expected from the simple ansatz employed in this work.
Nevertheless, even with the current ansatz, we expect that the FRG gives a better description than perturbative approaches for mixtures around the superfluid phase transition. Similarly, we find that the FRG gives a good description of repulsive Bose polarons at finite temperatures.

Having demonstrated that the FRG is capable of describing Bose-Bose mixtures, we devise several extensions of the current work that could study novel physics in quantum mixtures. The most straightforward extension is to study imbalanced Bose-Bose gases at zero and finite temperatures in both two and three dimensions. Even though the flow equations become much more complicated than with balanced mixtures, these should still be manageable for personal computers. Aside from the thermodynamics, it would be particularly interesting to study the phase-separation condition. Recent perturbative calculations have predicted that the phase separation depends strongly on the mass imbalance, and have also predicted novel mixed-bubble phases~\cite{naidon_mixed_2021}. Such physics could be studied with the current FRG formalism by examining the behaviour of the healing scales, which could provide a more robust picture than that from perturbative calculations. The study of the phase-separation condition at finite temperatures is also of interest, as it has been predicted that the temperature plays an important role~\cite{ota_magnetic_2019}.

Future studies should also consider better truncations with higher-order couplings. Standard higher-order couplings in the fields and momenta could be included to improve the accuracy of the presented results around the phase transition temperature. However, maybe more relevant is the inclusion of couplings that couple the phases of the two condensates in order to incorporate the effect of the superfluid drag. This could enable the FRG to accurately describe the superfluid phase transition in both two and three dimensions. In addition, in the case of polarons, it is relevant to include couplings that describe the feedback of the impurity onto the bath. This is particularly important to study heavy impurities.

The study of strongly-attractive Bose-Bose mixtures is of particular interest. The equation of state of Bose-Bose liquids, the onset of dimerisation and clustering, and the role of multi-body correlations are still open questions. The success of the FRG in the study of several resonant quantum gases makes it a promising tool to study these phenomena. In principle, strongly-attractive Bose-Bose mixtures could be studied with the FRG by simply proposing an ansatz with pairing fields as with Fermi gases~\cite{diehl_functional_2010}. However, the presence of the liquid phase could require more sophisticated ansatz\"e and initial conditions.

A more straightforward related extension is the study of resonant attractive Bose polarons at finite temperatures. This can be done by simply employing Matsubara sums over the approach presented in Ref.~\cite{isaule_renormalization-group_2021}. Several works have studied finite-temperature Bose polarons in the past few years with different approaches, sometimes leading to contradicting results. Therefore, FRG studies could help to provide a more definite answer. Moreover, it could be particularly interesting to study two-dimensional Bose polarons at finite temperatures in order to examine the impact of the BKT transition on the impurity.

In addition to homogeneous gases, the FRG has proved successful in studying lattice systems. In particular, it can give an accurate description of the quantum phases of the Bose-Hubbard model~\cite{rancon_nonperturbative_2011,rancon_nonperturbative_2011-1}, competing with MC simulations but with a much lower computational cost. The ansatz presented in this work can be extended to work on lattices by employing the lattice formulation of the FRG~\cite{machado_local_2010}. Such extension could provide robust descriptions of Bose-Bose mixtures and Bose polarons in optical lattices in two and three dimensions.

Finally, other related extensions include the study of coherently-coupled condensates~\cite{tommasini_bogoliubov_2003,abad_study_2013}, higher multi-component Bose gases, such as SU(N) Bose gases~\cite{hryhorchak_condensation_2020}, bipolarons~\cite{camacho-guardian_bipolarons_2018}, and impurities in superfluid Fermi gases~\cite{hu_crossover_2021,hryhorchak_polaron_2021}. Such systems can be studied with similar techniques to those used in this work.

\begin{acknowledgments}
We acknowledge useful discussions with B. Juli\'{a}-D\'{i}az and P. Massignan. F.I. acknowledges funding from EPSRC (UK) through Grant No. EP/V048449/1. I.M. acknowledges funding by Grant No. PID2020-114626GB-I00 from the MICIN/AEI/10.13039/501100011033. This work has been partially supported by MINECO (Spain) Grant No. FIS2017-87534-P. We acknowledge financial support from Secretaria d’Universitats i Recerca del  Departament  d’Empresa  i  Coneixement  de  la  Generalitat  de  Catalunya,  co-funded  by the European Union Regional Development Fund within the ERDF Operational Program of Catalunya (project QuantumCat, ref. 001-P-001644).
\end{acknowledgments}

\appendix

\section{The Wetterich equation}
\label{app:WettEq}

Here we present a short sketch for the derivation of the Wetterich equation ~(\ref{sec:FRG;sub:FRG;eq:WettEq}) for purely bosonic fields. For a more detailed derivation see Ref.~\cite{berges_non-perturbative_2002}.

First, we differentiate Eq.~(\ref{sec:FRG;sub:FRG;sec:Z}) with respect to $k$,
\begin{align}
    \partial_k W_k[\bm{J}] e^{W_k[\bm{J}]} =& -\frac{1}{2} \int\mathcal{D}\VPHI\left(\VPHI^\dag\partial_k\MR_k\VPHI\right)e^{-\mathcal{S}_k[\VPHI]+\int_x\bm{J}\cdot\VPHI}\nonumber\\
    =&-\frac{1}{2}\int_q\left(\delta_{\bm{J}^\dag}\partial_k\MR_k\delta_{\bm{J}}\right) e^{W_k[\bm{J}]}\nonumber\\
    =&-\frac{1}{2}\int_q\left(\PHI^\dag\partial_k\MR_k\PHI+\tr\left[\partial_k\MR_k\delta_{\bm{J}}\PHI\right]\right)e^{W_k[\bm{J}]}\,
\label{app:WettEq;eq:A1}
\end{align}
where $W_k=\ln\mathcal{Z}$ and $\PHI$ are the classical fields as defined in Eq.~(\ref{sec:FRG;sub:Gamma;eq:PHI}). Note that we have used Eq.~(\ref{sec:FRG;sub:FRG;eq:DSk}) in the first line. To continue we need an expression for both $\partial_k W_k$ and $\delta_{\bm{J}}\PHI$. By differentiating Eq.~(\ref{sec:FRG;sub:FRG;sec:Gammak}) with respect to $k$ we obtain
\begin{equation}
    \partial_k W_k = -\partial_k \Gamma_k[\PHI]-\partial_k \Delta\mathcal{S}_k[\PHI]=-\partial_k \Gamma_k[\PHI]-\frac{1}{2}\PHI^\dag\partial_k\MR_k\PHI\,,
\label{app:WettEq;eq:A2}
\end{equation}
whereas by taking two functional derivatives to Eq.~(\ref{sec:FRG;sub:FRG;sec:Gammak}) with respect to $\PHI$ we obtain
\begin{equation}
    \frac{\delta\bm{J}}{\delta\PHI}=\frac{\delta^2\Gamma_k[\PHI]}{\delta\PHI^\dag\delta\PHI}+\MR_k=\MGamma_k^{(2)}+\MR_k\,,
\label{app:WettEq;eq:A3}
\end{equation}
where we have once again used Eq.~(\ref{sec:FRG;sub:FRG;eq:DSk}) in both equations. By using the last two equations in (\ref{app:WettEq;eq:A1}) we obtain the Wetterich equation~(\ref{sec:FRG;sub:FRG;eq:WettEq}),
\begin{equation}
    \partial_k \Gamma_k = \frac{1}{2}\int_q\tr[\partial_k\MR_k(\MGamma_k^{(2)}+\MR_k)^{-1}]\,.
\end{equation}
Note that we have inverted Eq.~(\ref{app:WettEq;eq:A3}). Here we stress that for fermionic fields we need to consider Grassmann fields that anticommute. See Refs.~\cite{berges_non-perturbative_2002,dupuis_nonperturbative_2021} for more details.

\section{Driving terms}
\label{app:dkU}

The flow equation of the effective potential is simply obtained from Eq.~(\ref{sec:FRG;sub:FRG;eq:WettEq}). For the current truncation it reads
\begin{align}
    \partial_k U =&\frac{1}{2}\int_q\Big[\left(E_{A,2}(\Q)+V_A\omega_n^2\right)\left(E_{B,1}(\Q)+V_B\omega_n^2\right)\left(E_{B,2}(\Q)+V_B\omega_n^2\right)\nonumber\\
    &+\left(E_{B,2}(\Q)+V_B\omega_n^2\right)\left[\left(E_{a,1}(\Q)+V_A\omega_n^2\right)\left(E_{B,1}(\Q)+V_B\omega_n^2\right)-4\rho_A\rho_B\lambda_{AB}^2\right]\nonumber\\
    &+S_B^2\omega_n^2\left(E_{A,1}(\Q)+E_{A,2}(\Q)+2V_A\omega_n^2\right)\Big]\times\frac{\partial_k R_{k,A}(\Q)}{\det_{AB}(\Q)}+(A\leftrightarrow B)\,,
    \label{app:dkU;eq:dkU}
\end{align}
where
\begin{align}
    {\det}_{AB}(\Q)={\det}_{BA}(\Q)=&\left[S_A^2\omega_n^2+\left(E_{A,2}(\Q)+V_A\omega_n^2\right)\left(E_{A,2}(\Q)+V_A\omega_n^2\right)\right]\nonumber \\
    &\times\left[S_B^2\omega_n^2+\left(E_{B,2}(\Q)+V_B\omega_n^2\right)\left(E_{B,2}(\Q)+V_B\omega_n^2\right)\right]\nonumber\\
    &-4\rho_A\rho_B\left(E_{A,2}(\Q)+V_A\omega_n^2\right)\left(E_{B,2}(\Q)+V_B\omega_n^2\right)\lambda_{AB}^2\,,
\end{align}
and $E_{a,i}$ ($a=A,B$, $i=1,2$) are defined in Eqs.~(\ref{sec:FRGBB;sub:G;eq:Ea1}) and (\ref{sec:FRGBB;sub:G;eq:Ea2}). Note that these functions still depend on $\rho_a$. We stress that the momentum-integral is defined as in Eq.~(\ref{sec:model;eq:intq}).

The driving terms of Eqs.~(\ref{sec:FRGBB;sub:floweqs;eq:vkua}-\ref{sec:FRGBB;sub:floweqs;eq:vknpp}), that is, their RHS's, are simply obtained by differentiating (\ref{app:dkU;eq:dkU}) with respect to $\rho_a$ and $\tilde{\mu}_a$ and then evaluating at $\rho_a=\rho_{a,0}$ and $\tilde{\mu}_a=\mu_a$ for $a=A,B$.

The driving terms for the two-point functions are dictated by the diagrams in Fig.~\ref{sec:FRGBB;sub:floweqs;fig:Gamma2}. These are given by
\begin{multline}
    \partial_k \Gamma^{(2)}_{a_i,b_j}=\int_q\tr\bigg[\partial_k \MR_k(\Q)\MG_k(q)\MGamma^{(3)}_{b_j}(-q,p,q+p)\MG_k(q+p)\MGamma^{(3)}_{a_i}(-q-p,p,q)\MG_k(q)\nonumber\\
    -\frac{1}{2}\partial_k \MR_k(\Q)\MG_k(q)\MGamma^{(4)}_{b_j,a_i}(-q,q,p,-p)\MG_k(q+p)\bigg]\,,
    \label{app:dkU;eq:dkGamma2}
\end{multline}
where in our current truncation the three-point vertices are given by
\begin{align}
    \MGamma^{(3)}_{a_1}=&\begin{pmatrix}
    3 (2\rho_a)^{1/2}\lambda_{aa} & 0 & (2\rho_b)^{1/2}\lambda_{AB} & 0 \\
    0 & (2\rho_a)^{1/2}\lambda_{aa} & 0 & 0 \\
    (2\rho_b)^{1/2}\lambda_{AB} & 0 & (2\rho_a)^{1/2}\lambda_{AB} & 0 \\
    0 & 0 & 0 & (2\rho_a)^{1/2}\lambda_{AB}
    \end{pmatrix}\,,& a\neq b \\
     \MGamma^{(3)}_{a_2}=&\begin{pmatrix}
    0 & (2\rho_a)^{1/2}\lambda_{aa} & 0 & 0 \\
    (2\rho_a)^{1/2}\lambda_{aa} & 0 & (2\rho_b)^{1/2}\lambda_{AB} & 0 \\
    0 & (2\rho_b)^{1/2}\lambda_{AB} & 0 & 0 \\
    0 & 0 & 0 & 0
    \end{pmatrix}\,.& a\neq b
\end{align}
Within the truncation employed in this work all vertices are momentum-independent, and thus, the second diagram in Fig.~\ref{sec:FRGBB;sub:floweqs;fig:Gamma2} (the term in the second line of Eq.~(\ref{app:dkU;eq:dkGamma2})) does not depend on the external momentum $p$. Therefore, this term does not contribute to the flows of $S_a$, $Z_a$, and $V_a$, and so we do not need the four-point functions. Nevertheless, we refer to Refs.~\cite{berges_non-perturbative_2002,dupuis_nonperturbative_2021} for more discussion on the hierarchy of the FRG flow equations.

To generate the driving terms for this or other FRG studies it is convenient to employ a symbolic programming language. In particular, we refer the reader to the \texttt{DoFun} package~\cite{huber_dofun_2020}, which automatically generates FRG flow equations for a given ansatz. Alternatively, the reader can employ standard packages for functional differentiation. 

\bibliography{biblio}

\end{document}